\documentclass[]{aa}

\usepackage[dvips]{graphicx}
\usepackage{lscape,aalongtable}
\usepackage{supertabular}
\usepackage{amsmath}
\usepackage{amssymb}

\newcommand{\Msol}{\mbox{\rm M$_{\odot}$}}

\newcommand{\Lsol}{\mbox{\rm L$_{\odot}$}}

\usepackage[]{natbib}
\bibpunct{(}{)}{;}{a}{}{,}
\begin{document}


\title{IRAS 18511+0146: a proto Herbig Ae/Be cluster?
}

\author{
S. Vig$^1$, L. Testi$^{1,2}$, M. Walmsley$^1$, S. Molinari$^3$, S. Carey$^4$, 
A. Noriega-Crespo$^4$
}

\offprints{S. Vig, \email{sarita@arcetri.astro.it}}

\institute{$^1$INAF-Osservatorio Astrofisico di Arcetri, Largo E. Fermi 5, I-50125 Firenze, Italy\\
$^2$ESO, Karl Schwarzschild str. 2, D-85748 Garching, Germany \\
$^3$INAF-Istituto di Fisica dello Spazio Interplanetario, Via Fosso del 
Cavaliere, I-00133 Roma, Italy\\
$^4$Spitzer Science Center, California Institute of Technology, Pasadena, CA 
91125, USA \\ 
}


\abstract{The evolution of a young protocluster depends on the relative 
spatial distribution and dynamics of both stars and gas.}
{We study the distribution and properties of the gas and stars surrounding 
the luminous ($10^4$ \Lsol) protocluster IRAS 18511+0146.}
{IRAS 18511+0146 and the cluster associated with it has been 
investigated using the sub-millimetre (JCMT-SCUBA), infrared (Spitzer-MIPSGAL, 
Spitzer-GLIMPSE, Palomar) and radio (VLA) continuum data. Cluster 
simulations have 
been carried out in order to understand the properties of clusters as well as 
to compare with the observations.}
{The central most obscured part of the protocluster coincident with the 
compact sub-millimetre source found with SCUBA is responsible for at least 2/3 
of the total luminosity. A number of cluster members have been identified which 
are bright in mid infrared and show rising (near to mid infrared) 
spectral energy distributions 
suggesting that these are very young stellar sources. In the mid infrared 
8.0~$\mu$m image, a number of filamentary structures and clumps are detected in 
the vicinity of IRAS 18511+0146. }
{Based on the luminosity and cluster size as well as on the
evolutionary stages of the cluster members, IRAS 18511+0146 is likely 
to be protocluster with the most massive object being a precursor to a Herbig 
type star.}


\titlerunning{IRAS 18511+0146: a proto Herbig Ae/Be cluster?}
\authorrunning{S. Vig, et al.}
\maketitle


\section{Introduction}

Massive stars form accompanied by swarms of lower mass objects and the 
relationship between the two is by no means clear. It is however known  that 
the development of a young cluster depends sensitively on events which occur 
in the phase when it is still surrounded by the remnants of the ``core" from 
which it formed \citep{2003ARA&A..41...57L}. It is therefore of 
interest to obtain sensitive observations 
of the young protostars and the surrounding gas in the phase when dust 
obscuration causes the cluster to be invisible at optical wavelengths.  
The objects of interest are thus best observed at infrared and radio 
wavelengths. 

This article reports such a study of the cluster surrounding what appears to 
be a young intermediate (2-8 \Msol) protostar (IRAS 18511+0146, 
also known as Mol~75 and RAFGL 5542) from the survey of 
\citet{1996A&A...308..573M,1998A&A...336..339M} using both data from the 
\textit{Spitzer} satellite and ground based observations. We proceed on 
the hypothesis that the cluster surrounding IRAS 18511+0146 is essentially a 
forerunner of the small clusters examined by \citet{1998A&AS..133...81T}. 
Thus, the bolometric luminosity (roughly $10^4$ \Lsol) and size (0.5 parsec) 
are similar but, as discussed later, the visual extinction is large (of order 
50 visual magnitudes) and variable.

The gas distribution is of great importance both because of its effect upon 
the extinction to individual protostars and because the gas plays an 
important role dynamically. The outflows and ionization caused by the young 
protostars are thought to be the primary agent causing the cluster to 
disperse. We therefore use JCMT-SCUBA observations to assess the mass and 
distribution of the gas associated with IRAS 18511+0146. We then use this 
information in  simulations which we have carried out with the aim of testing 
our hypothesis that the young protocluster associated with IRAS 18511+0146 is 
an obscured version of the intermediate mass clusters studied by 
\citet{1998A&AS..133...81T}.

Previous studies of IRAS 18511 were carried out by \citet{1999ApJS..125..143W}
 using the OVRO interferometer and the VLA to map both molecular line and 
continuum emission. They detected what appeared to be a small ionized  region 
ionized by a B1 star which coincided with a compact clump seen in C$^{18}$O. 
\citet{2004ApJS..155..149K} detected a methanol maser which was however 
offset by roughly 0.4 pc ($\sim$19\arcsec) relative to the ionized gas for 
a distance of 3.9 kpc to IRAS 18511 \citep{1996A&A...308..573M}. 
\citet{2002AJ....124.2790I} detected the 3~$\mu$m water ice feature  
towards the central object suggesting a visual 
extinction of 22 magnitudes along this line of sight. 
\citet{2005ApJ...625..864Z} have reported strong indications of an 
outflow in IRAS 18511 using the $^{12}$CO line. According to them, the 
mass-loss rate is $\sim$3.7$\times10^{-4}$ \Msol\ 
yr$^{-1}$ and the dynamical timescale is 5$\times10^4$ yrs. 
\citet{2001A&A...370..230B} imaged IRAS 18511 using the IRAM 30-m and 
KOSMA 
telescopes in a number of molecular lines ($^{13}$CO, CS, HCO$^+$) and find 
blue shifted and red shifted components distributed over various clumps.

In this study, we explore in detail the region surrounding the 
IRAS source 18511+0146 using infrared, sub-millimetre, and radio 
continuum data. In Sect. 2, we summarize the observations. 
In Sect. 3, we present the observational results and in 
Sect. 4, a model of the 
relative distributions of gas and stars is examined. A discussion of IRAS 
18511 region is presented in Sect. 5, and a short summary 
of our conclusions is given in Sect. 6.

\section{Observations, Available data and Data reduction}

\subsection{JCMT-SCUBA}

The sub-millimetre observations of IRAS 18511 at 450 and 850~$\mu$m, using the
Submillimetre Common User Bolometer Array (SCUBA) of the James Clerk Maxwell 
Telescope\footnote{This paper makes use of data from the James Clerk Maxwell 
Telescope (JCMT) Archive. The
JCMT is operated by the Joint Astronomy Centre on behalf of the UK Particle
Physics and Astronomy Research Council, the National Research Council of
Canada and the Netherlands Organisation for Pure Research.}
 were carried out on 28 May 1999. The data were processed using their 
standard pipeline SCUBA User Reduction
 Facility (SURF). The planet Uranus was used for calibration. Submillimetre
maps were generated at 450 and 850~$\mu$m and the fluxes extracted. The
beam size is 10\arcsec\ at 450~$\mu$m and 15\farcs5 at 850~$\mu$m. The
sensitivities are 0.3 Jy/beam and 0.04 Jy/beam at 450 and 850~$\mu$m,
respectively. The flux densities extracted were used to construct the Spectral 
Energy Distribution (SED) of IRAS 18511.

\subsection{Spitzer data}

The Spitzer Space Telescope\footnote{This
work is based in part on observations made with the
\textit{Spitzer Space Telescope}, which is operated by the Jet Propulsion
Laboratory, under NASA contract 1407.} \citep{2004ApJS..154....1W} was launched
in space in August 2003 and consists of a 0.85-meter telescope with three 
cryogenically cooled instruments: InfraRed Array Camera (IRAC), Multiband 
Imaging Photometer for Spitzer (MIPS) and InfraRed Spectrograph. We have used 
data from the Legacy projects, GLIMPSE (PI: E. Churchwell) and MIPSGAL 
(PI: S. Carey) in this paper. The GLIMPSE and MIPSGAL images have been obtained
using the software `Leopard'.

\subsubsection{MIPSGAL}

The Multiband Imaging Photometer for Spitzer (MIPS) provides the Spitzer Space 
Telescope with capabilities for imaging and photometry in broad spectral bands 
centred nominally at 24, 70, and 160~$\mu$m, and for low-resolution 
spectroscopy between 55 and 95~$\mu$m \citep{2004ApJS..154...25R}. The 
Multiband Imaging Photometer for Spitzer Galactic Planey survey (MIPSGAL) 
surveyed the sky in identical regions as covered by GLIMPSE (next subsection) 
at 24 and 70~$\mu$m using the MIPS instrument \citep{2005AAS...207.6333C}. The 
instrument achieves diffraction-limited resolution of 6\arcsec\ and 18\arcsec\ 
at 24 and 70~$\mu$m, respectively. The pixel size is 2\farcs55 at 24~$\mu$m 
and 9\farcs98 at 70~$\mu$m. The Spitzer Science Center provides final mosaics 
called Post-Basic Calibrated Data (PBCD) products. These are the 
maps of 
multiple calibrated image frames or BCDs  (Basic Calibrated Data).  For 
IRAS 18511, we found that these products satisfied our scientific goals.
We have carried out extraction and photometry of the sources in the region 
around IRAS 18511 using the MIPSGAL PBCD images at 24~$\mu$m and 70~$\mu$m 
(details given in Appendix A). The details of the extracted sources and their 
fluxes are presented in Section 3.

\subsubsection{GLIMPSE}

In the GLIMPSE \citep[Galactic Legacy Infrared Midplane Survey 
Extraordinaire;][]{2003PASP..115..953B} project, the Spitzer Space Telescope
surveyed approximately 220
square degrees of the Galactic plane covering a latitude range of
$|b|<1^{\circ}$ and a longitude range of $10^{\circ}\le l\le65^{\circ}$,
$-65^{\circ}\le l\le-10^{\circ}$. This survey is carried out in the 4
IRAC bands. IRAC is a four-channel camera that provides
simultaneous 5\farcm2$\times$5\farcm2 images at 3.6, 4.5, 5.8, and 8.0~$\mu$m 
with a pixel size of 1\farcs2$\times$1\farcs2  \citep{2004ApJS..154...10F}.
The resolutions achieved by IRAC are 2\farcs4, 2\farcs4, 2\farcs8 and 3\farcs0 
in the 3.6, 4.5, 5.8 and 8.0~$\mu$m bands, respectively.

The sources in the IRAS 18511 region have been extracted from the GLIMPSE More 
Complete Archive. The GLIMPSE archive
contains point sources with peak signal-to-noise ratio greater than 5
in at least one band. The extracted sources have been used in constructing the 
spectral energy distributions. However, the bright/saturated sources are not 
extracted in the GLIMPSE-catalog. In order to get a lower limit on the 
fluxes of these sources, we have carried out aperture photometry
of these bright/saturated sources on GLIMPSE-IRAC PBCD images. For these 
bright sources, we have taken an aperture radius of 10 pixels and a 
sky annulus of 15 pixels in order to estimate the fluxes. The approximate 
centre was determined from the radial profiles using the task `imexamine' in 
Image Reduction and Analysis Facility (IRAF). For such a combination of 
aperture radius and sky annulus, no aperture correction needs to be applied. 

The PBCD images have been used to study the spatial distribution of sources as 
well as near and mid infrared emission from this region.

\subsection{Palomar Data}

Deep near infrared observations of a $\sim$3\arcmin$\times$3\arcmin field
surrounding IRAS 18511 were obtained
on 26 July 1999 with the Palomar Observatory 60-inch telescope equipped with
the near infrared IRC-NICMOS3 camera. The field was observed in three broad 
bands: J, H, and K$_s$. A standard dithering technique was used to
efficiently remove the sky emission and correct for hot or dead pixels in
the 256$\times$256 NICMOS3 detector. The data were processed using 
recipes in IRAF to produce the final images.
Photometric calibration was achieved by observing a set of standard
stars from the lists of \citet{1998AJ....115.2594H} and
\citet{1998AJ....116.2475P}.
Astrometric calibration was performed tying the observed positions of
bright and isolated sources in the Palomar field to the
corresponding entries in the Two Micron All-Sky Survey (2MASS) database. We 
estimate this procedure to be accurate within 0\farcs5.

The limiting magnitudes of our observations
are found to be 18.1, 18.0, and 16.9 in the J, H, and K$_s$ bands, respectively.

\subsection{VLA}

We have also analysed the data for IRAS 18511 at 8.5 and 15 GHz taken from the
NRAO Data Archive (project ID: AD406) obtained using the Very Large Array
(VLA) in the D configuration. The observation was carried out in Dec 1997 by
\citet{1999ApJS..125..143W}. The total on-source integration time in 
each band is 30 min. The flux calibrators were 1331+305 and 0137+331. 
1832-105 was used as the phase calibrator. The NRAO Astronomical Image 
Processing System (AIPS) was used for reduction of the data. The beam
sizes are 8\farcs5$\times$7\farcs2 and 5\farcs2$\times$4\farcs4 at 8.5 and 15
GHz, respectively. The rms noise in the maps
are 0.05 mJy/beam and 0.13 mJy/beam at 8.5 and 15 GHz, respectively.

\section{Results}

\subsection{Submillimetre emission from cold dust}

\begin {figure*}
\includegraphics[height=9.0cm,angle=-90]{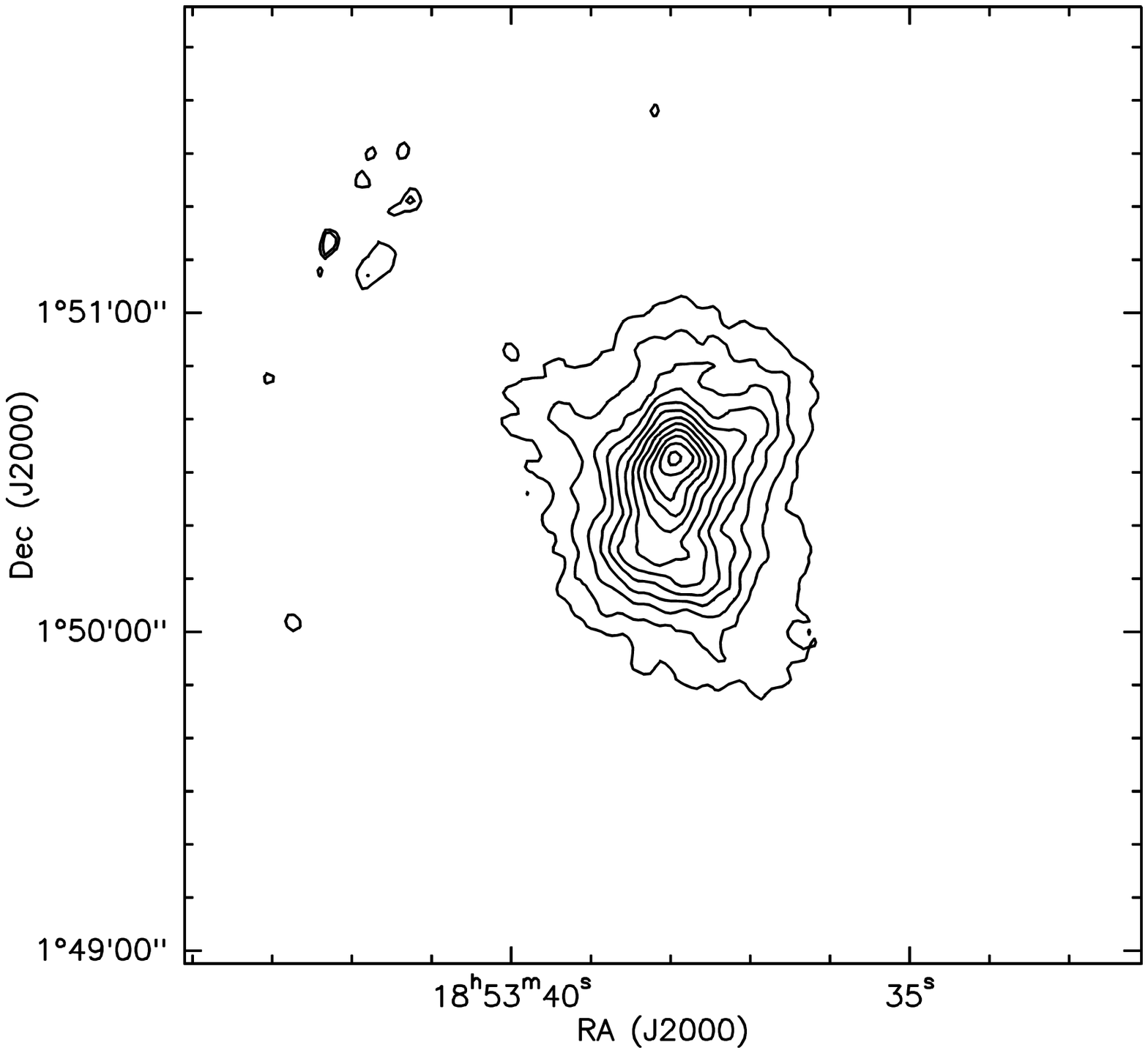}
\includegraphics[height=9.0cm,angle=-90]{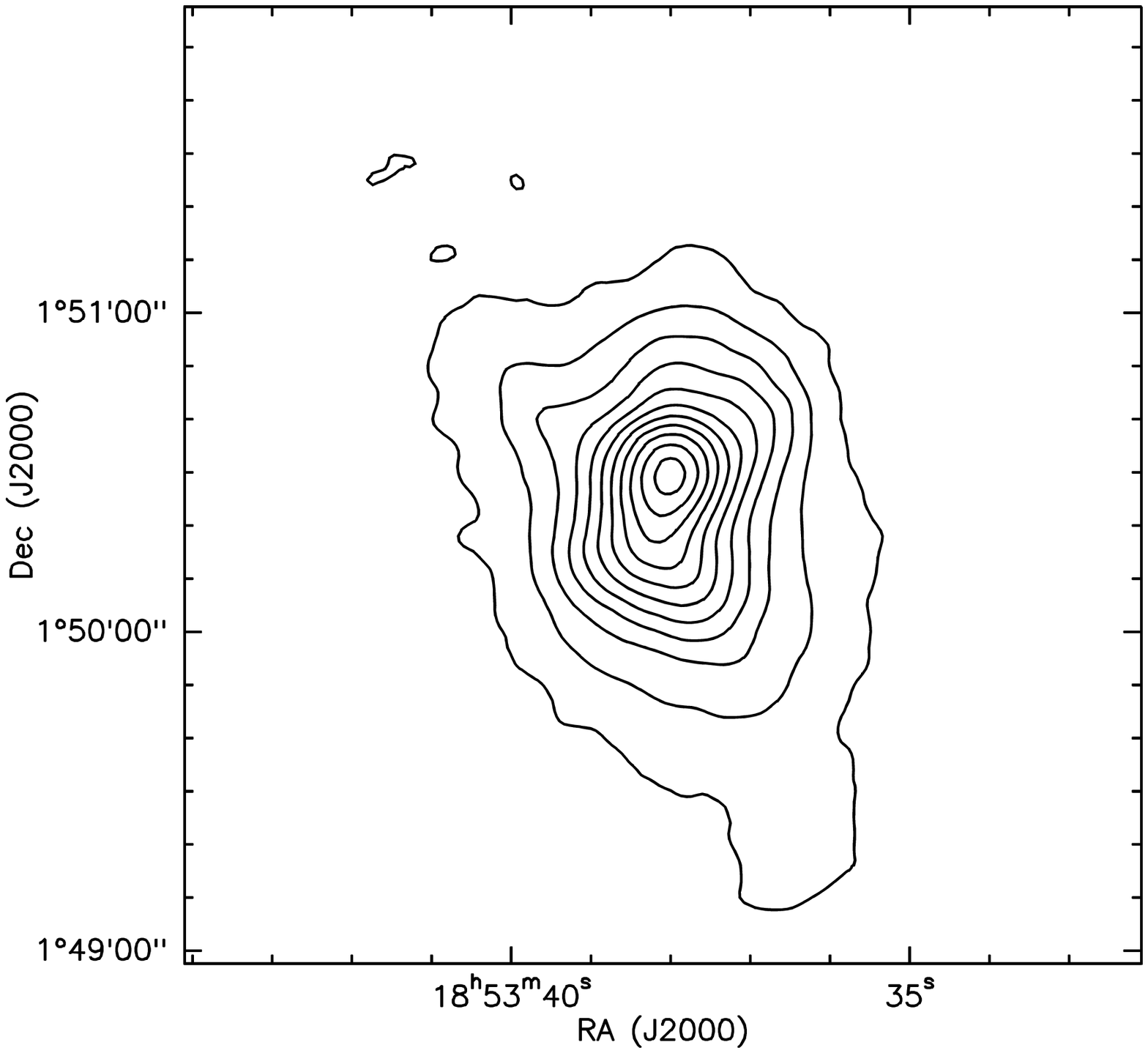}
\caption{(Left) JCMT-SCUBA 450~$\mu$m emission from IRAS 18511 region. The 
contours levels are from 2.0 to 13.4 Jy/beam (peak flux density 
$\sim$ 13.4 Jy/beam) in steps of 1 Jy/beam, where the beam size is 10\arcsec. 
(Right) JCMT-SCUBA 850~$\mu$m emission from IRAS 18511 region. The contours 
levels are from 0.18 to 2.28 Jy/beam (peak flux density $\sim$ 2.28 
Jy/beam) in steps of 0.2 Jy/beam, beam is 15\farcs5. 
}
\label{jcmt}
\end {figure*}

The JCMT-SCUBA maps at 450 and 850~$\mu$m trace the emission from cold dust
in the region of IRAS 18511. The JCMT maps are shown in Fig. \ref{jcmt}. The
integrated flux densities upto 10$\%$ contour levels are 139 Jy and 14 Jy at 
450 and 850~$\mu$m, respectively. The peak flux densities are 1.33 Jy/beam and 
2.28 Jy/beam at 450 and 850~$\mu$m, respectively. The half-power sizes  
are $0.4\times0.6$ pc$^2$ at 450~$\mu$m and 
$0.5\times0.7$ pc$^2$ at 850~$\mu$m. The emission maps show the 
presence of a core ($\alpha_{J2000}$, $\delta_{J2000}$ $\sim$ 18$^h$ 53$^m$ 
38\fs00, +01\degr 50\arcmin 29\farcs1) as well as extended emission, 
particularly to the south. The cold dust emission shows an extension 
towards the south-east ($\alpha_{J2000}$, $\delta_{J2000}$ $\sim$ 18$^h$ 
53$^m$ 38\fs13, +01\degr 50\arcmin 21\farcs6) which veers towards west 
further south. Emission towards the north-east of the core is also 
discerned.

The emission at 450~$\mu$m as well as at 850~$\mu$m can be used to estimate the
dust temperature of the core as follows. We know that the flux density, 
$F_\nu$, for
optically thin emission can be written as
\begin{equation}
F_\nu = \Omega B_\nu(T_d)\tau_\nu
\end{equation}
where $\Omega$ is the beam solid angle, $B_\nu$ is the Planck function, 
$\tau_\nu$ is the optical depth at 
frequency $\nu$ and $T_d$ is the dust temperature. Assuming, $\tau_\nu 
\propto \nu^\beta$, the ratio of flux densities at any 
two wavelengths is a function of $T_d$ and $\beta$. We have assumed $\beta=2$
and used the ratio of fluxes (after convolving the 450~$\mu$m map to the beam
resolution of 15\farcs5 corresponding to the 850~$\mu$m beam) in order to 
obtain 
a dust temperature of $37\pm5$ K at the peak of the emission. The 
total mass of the cloud has been 
estimated using the formalism of \citet{1983QJRAS..24..267H}. Assuming an 
average dust temperature of $20-30$ K for the entire cloud and using the 
value of dust emissivity at 850~$\mu$m ($\kappa_{850 \mu m} = 1.5$ cm$^2$ 
g$^{-1}$) appropriate to dark clouds from
\citet{2003A&A...399L..43B}, the total mass of the cloud is estimated to be
$\sim750-1310$ \Msol. This mass estimate can be compared with 
the gas mass of 826 \Msol\ obtained by \citet{1999ApJS..125..143W} using the 
C$^{18}$O measurements.
 Assuming 750 \Msol to be distributed in a spherical volume 
of radius 0.6 pc, we find the mean density to be $3.4\times10^4$ cm$^{-3}$. 
For this mass and mean density, the free fall time of the cloud is estimated 
to be $\sim3\times10^5$ yrs.  

The visual extinction (in magnitudes) for constant temperature and dust
density distribution along the line-of-sight can be obtained by using the
following expression:
\begin{equation}
I_\nu = \frac{\kappa_\nu}{\kappa_V} \times \frac{A_V}{1.086} \times B_\nu(T)
\end{equation}
where $I_\nu$ is the flux density along the line-of-sight, $\kappa_\nu / \kappa_V$ is the dust emissivity normalised to emissivity in the  V-band, and
$A_V$ is the extinction along the line-of-sight.
For a dust temperature of $37\pm5$ K at IRAS 18511 core and 
$\kappa_{850 \mu m} / \kappa_V = 4.0\times10^{-5}$ obtained by 
\citet{2003A&A...399L..43B} for dark clouds, the maximum extinction 
using the peak 850~$\mu$m flux is estimated to be $86\pm15$ 
magnitudes. The average 
extinction has been estimated assuming that the total mass is distributed 
uniformly in the area enclosed within the 10\% contour level  
($\sim1.2\times1.8$ pc$^2$) at 850~$\mu$m. Assuming a temperature of 
$20-30$ K over this region, the average value of extinction is 
estimated to be $A_V\sim42-24$ magnitudes. 

\subsection{Mid infrared emission}

\begin {figure*}
\includegraphics[height=9.0cm,angle=-90]{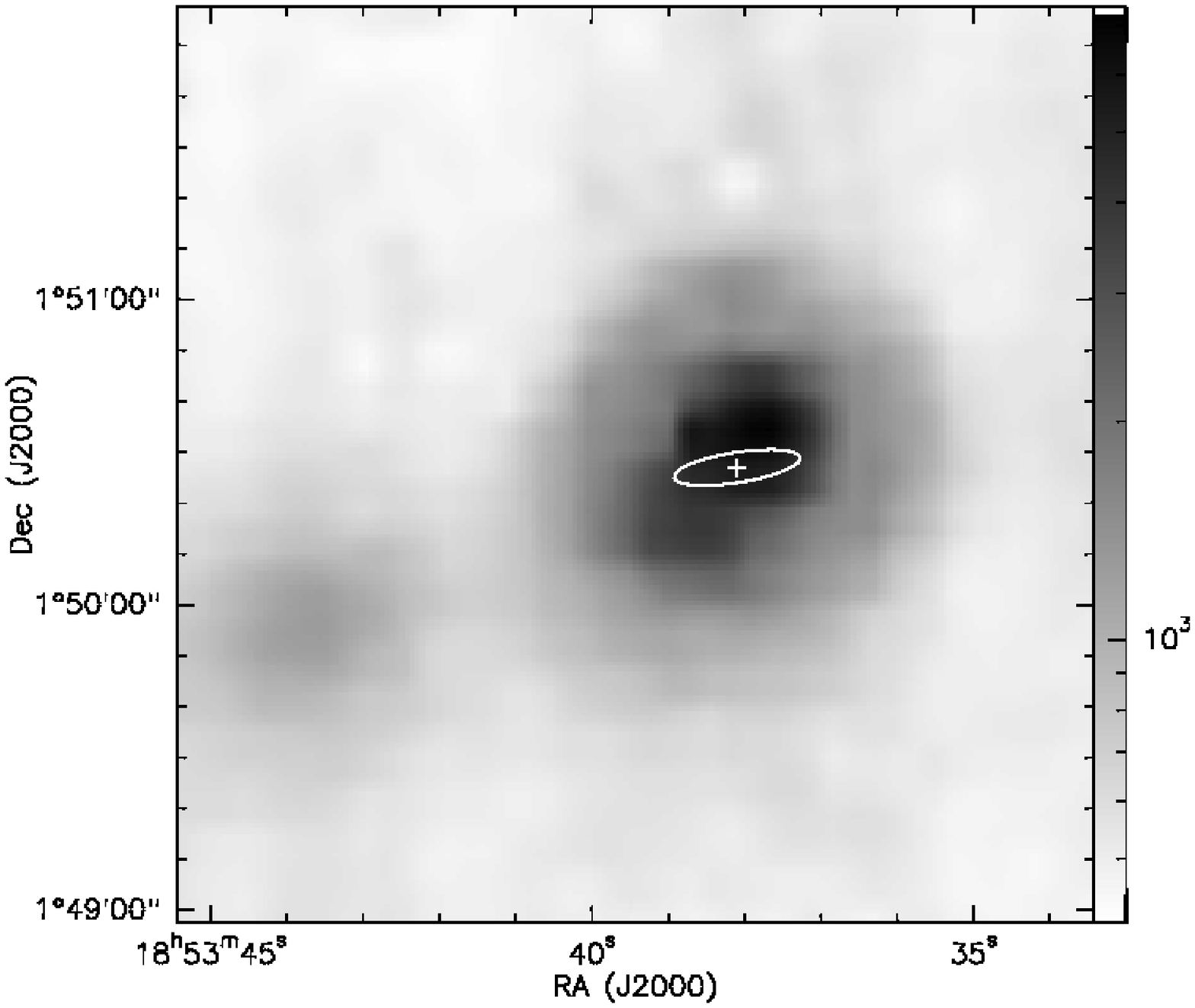}
\includegraphics[height=9.0cm,angle=-90]{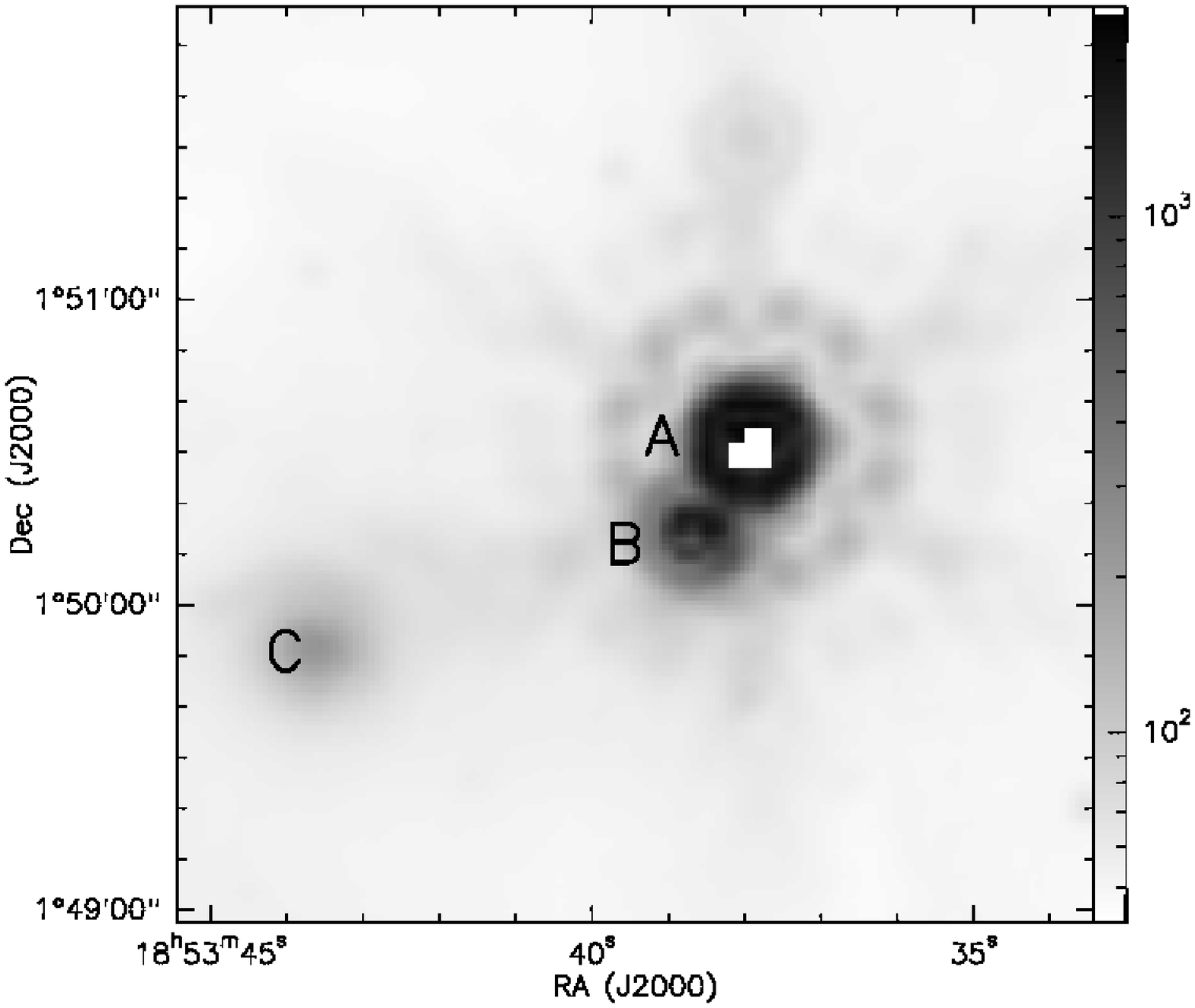}
\includegraphics[height=9.0cm,angle=-90]{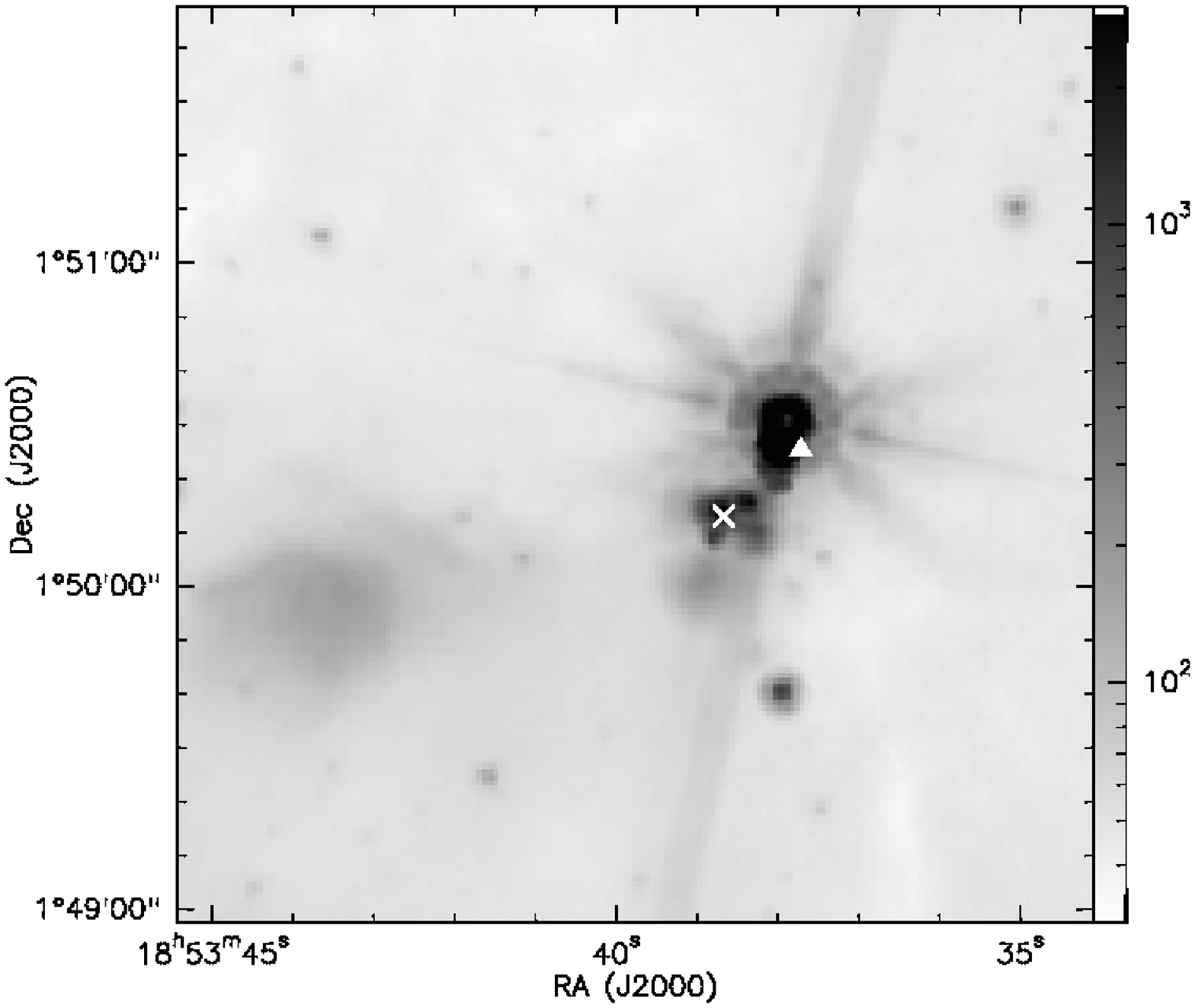}
\includegraphics[height=9.0cm,angle=-90]{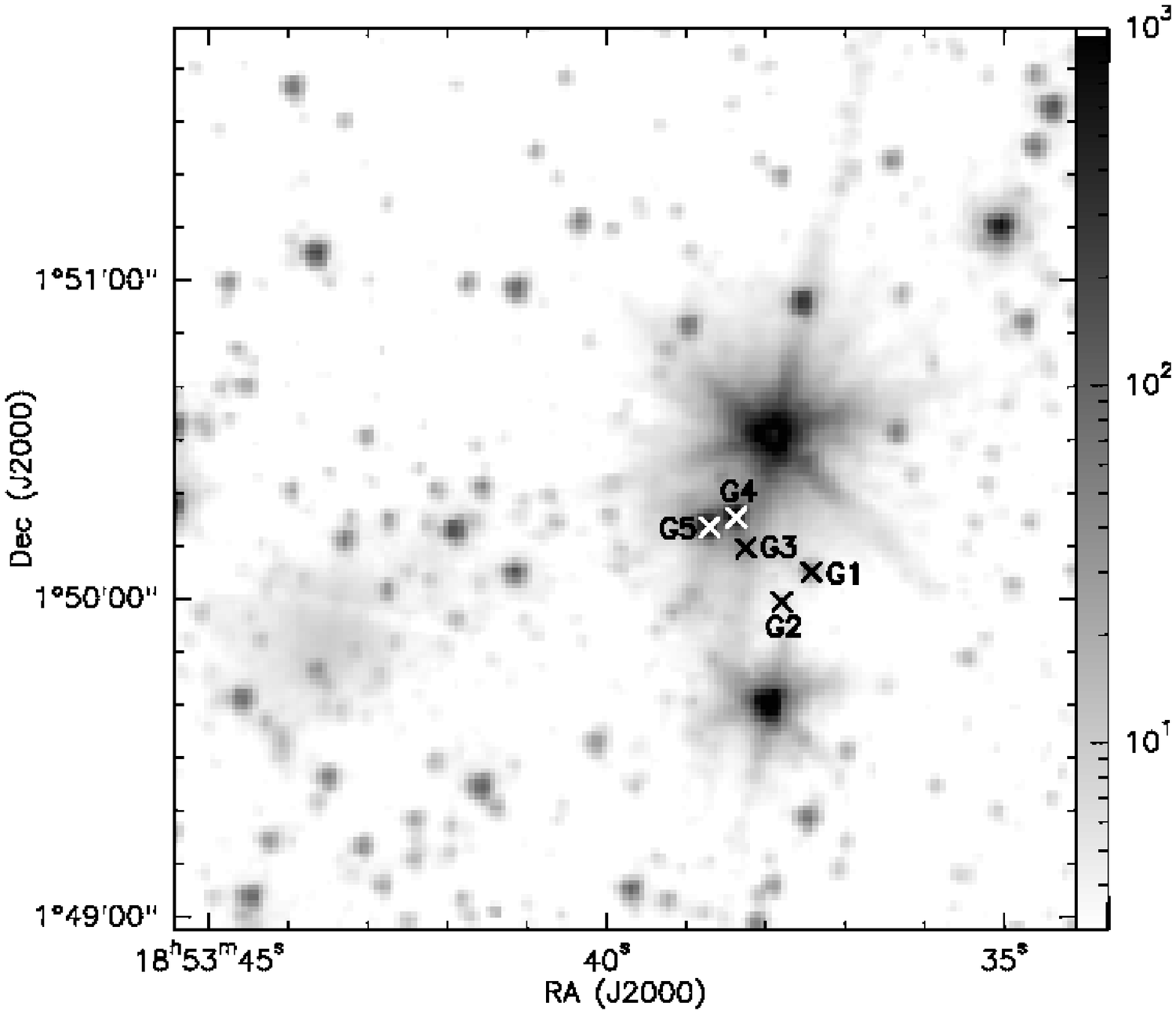}
\caption{(Top left)MIPS 70~$\mu$m image of IRAS 18511 region. The plus 
symbol and the ellipse represent the IRAS position and the corresponding 
1-$\sigma$ position error ellipse. (Top right) MIPS 
24~$\mu$m image of IRAS 18511 region with the sources A, B and C marked. 
(Bottom left) IRAC 8.0~$\mu$m image of IRAS 18511 region. The white cross marks 
the position of radio emission at 8.5 and 
15 GHz. The solid triangle marks the position of the methanol maser.
(Bottom right) IRAC 3.6~$\mu$m image with sources G1-G5 marked. 
}
\label{glim_mips}
\end {figure*}

The MIPSGAL 70~$\mu$m image shows the presence of three sources in this region.
This image is shown in Fig. \ref{glim_mips}. The IRAS position is also 
marked. These sources 
are better resolved in the MIPS 24~$\mu$m image. For convenience, we call 
these sources A, B and C. The location of these sources is shown on the 
24~$\mu$m image in Fig. \ref{glim_mips}. While the IRAS peak coinicides with A,
 B lies nearly 20\arcsec\ (0.4 pc) to the south-west of A. C 
is the faintest of the three and is separated by $\sim$90\arcsec\ from A. A is 
saturated in the MIPS 24~$\mu$m image. The fluxes and positions of these 
sources 
have been extracted using the methods described in the Sect. 2.2.1. 
Table ~\ref{fluxes} lists the positions as well as fluxes of A, B and C, 
respectively.

The four GLIMPSE-IRAC images show a bright source at A and a small compact
group of stars located to the south-west of this source (associated with B).
Figure \ref{glim_mips} shows the locations of A, B and C on the grayscale 
8.0~$\mu$m image. The IRAC 3.6~$\mu$m image of IRAS 18511 is also shown in Fig. 
\ref{glim_mips}. The bright source associated with A is saturated in all the 
four IRAC bands. However we have extracted the lower limits to the flux of 
this source using the method of aperture photometry as mentioned in Sect. 
2.2.2. A comparison of the MIPSGAL 24~$\mu$m image with the 
higher angular resolution IRAC image shows that B consists of a group
of sources. One can see an extension corresponding to A in the 
8.0~$\mu$m image suggesting that A may consist of two or more sources. 
In the GLIMPSE 3.6 or 4.5~$\mu$m images, the emission from C is very 
faint. 
The emission from C is diffuse in appearance and stronger in the 
8.0~$\mu$m image.

\subsubsection{The filaments around IRAS 18511}

\begin {figure*}
\includegraphics[height=9.0cm,angle=-90]{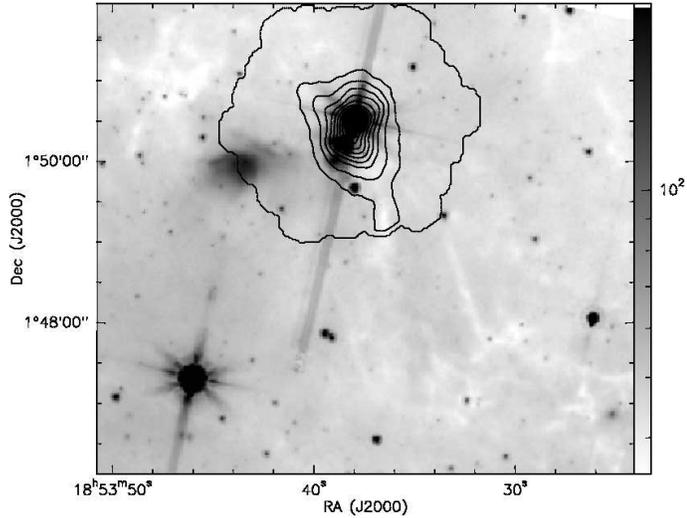}
\caption{Grayscale image of IRAS 18511 at GLIMPSE-IRAC 8.0~$\mu$m showing the dark
filaments against diffuse emission. Overlaid as contours is the JCMT-SCUBA
sub-millimetre emission at 850~$\mu$m. The region covered by the SCUBA 
map is also shown.
}
\label{IRD}
\end {figure*}

An investigation into the 8.0~$\mu$m image shows diffuse emission in
the neighbourhood of IRAS 18511. It is particularly interesting to note the
filamentary structures (white in Fig. \ref{IRD}) seen in absorption 
against this diffuse emission. IRAS 18511 seems to lie on one clump and 
extension of this clump (in the form of filaments)
can be seen towards the south-west of A. A clumpy structure can be observed
towards the east of IRAS 18511.  An estimate of extinction 
towards these filamentary and clumpy structures has been obtained 
by using the following relation:
\begin{equation} 
I = I_o\; e^{-\tau_8} 
\end{equation}
where $I_o$ is the background source intensity (determined in the 
vicinity of the filamentary structures), $I$ is the observed 
intensity, and $\tau_8$ is the optical depth at 8.0~$\mu$m. One observes 
considerable variations of $I_o$ across the image.
We have therefore assumed a different value of $I_o$ for each pixel as 
follows. A median filter image has been constructed using 
a window size of 99 pixels (2\arcmin). Using the median value at every pixel 
as the corresponding $I_o$, a visual extinction (A$_V$) map has been 
constructed taking $A_V$/$\tau_8=22.5$. This relation ($A_V$/$\tau_8$) has 
been obtained by taking the
ratios of extinctions ($A_8$/$A_K=0.43$, $A_K$/$A_V=0.112$) obtained by
\citet{2005ApJ...619..931I} and \citet{1985ApJ...288..618R}.  The 
pixels representing values higher than the median have been masked. The
A$_V$ values are found to have values in the range $6-8$ in the filaments as
well as towards the clumpy structures. Converting the A$_V$ map to 
a column density map, the mass in the filaments as well as in 
the clumps have been estimated. While the mass
 of gas in the clump ($\alpha_{J2000}, \delta_{J2000} \sim$ $18^h$ $53^m$
47\fs3 +01\degr 51\arcmin 0\arcsec) is found to be $\sim46$ \Msol, the mass
in a few filaments is found to be in the range $\sim5-15$ \Msol. 
 The total mass of the filaments and
clumpy structures shown in Fig. \ref{IRD} is $\sim120$ \Msol. These mass 
estimates represent lower limits as $I_o$ has contributions from 
foreground as well as background emission. In addition, 
the total mass estimate also suffers from blanking of pixels near the
8.0~$\mu$m sources. In particular, since IRAS 18511 A is very bright at 8.0~$\mu$m,
the clump containing IRAS 18511 A is masked out in the
extinction map. It is interesting to note that there is an overlap of a 
filament in the 8.0~$\mu$m image with the JCMT-SCUBA 850~$\mu$m sub-millimetre 
emission towards the south-west.

\subsubsection{Associated young stellar objects (IRAC Colour-colour diagram)}

In order to select IRAC sources from the GLIMPSE catalog for further study, 
we have defined a `region of interest' around IRAS 18511. We take this to be a 
region overlapping the sub-millimetre emission from cold dust. In other words, 
we take the sources in a region 
enclosed by the 10\% contour level of the peak of the the sub-millimetre 850~$\mu$m emission. We find a total of 39 sources (including the saturated 
source at A). Of these, six are detected in all four IRAC bands. These
have been plotted in the IRAC colour-colour diagram ([3.6]-[4.5] vs.
[5.8]-[8.0]) which is shown in Fig. \ref{ccd}. In the diagram, the solid 
square approximately delineates the region occupied by Class II sources 
whereas the dotted square covers the region occupied by the Class I 
models of \citet{2004ApJS..154..363A} (see their Fig. 4). From the 
colour-colour diagram in Fig. \ref{ccd}, we find five sources lying within the 
boxes representing the region occupied by either Class I or 
Class II sources. 
We label these sources G1, G2, G3, G4 and G5. Of these 5 objects, G2, G3, 
G4 and G5 lie within the Class I box while G1 lies in the overlap region 
between Class I and Class II. These sources are also
shown in Fig. \ref{glim_mips} and the details of the coordinates and fluxes of 
these sources are included in Table \ref{ysolist}. 

\subsection{Near infrared emission}

The near infrared Palomar K$_s$ band image of IRAS 18511 is 
shown in Fig. \ref{Pal}.  Unlike many other star forming regions, we do not 
detect near infrared nebulosity in these images. 
Among the sources extracted from the Palomar images in the J, H and K$_s$
 bands, we select a sample of sources lying within the `region of interest'
(see Sect. 3.2.2 for details). We find a total of 68 sources, including the 
saturated sources. The fluxes of three sources which are saturated in the 
Palomar images have been taken from the 2MASS\footnote{This publication makes use of data products from the Two Micron All Sky
Survey, which is a joint project of the University of Massachusetts and the
Infrared Processing and Analysis Center/California Institute of Technology,
funded by the NASA and the NSF.}. 
This includes the source associated with A, which is saturated in the H and 
K$_s$ bands of the Palomar images. Of the total of 68 sources, we find that 
27 sources are detected in all three JHK$_s$ bands.

\subsubsection{Associated young stellar objects (NIR and IRAC colour-colour 
diagrams)}

In order to identify the young stellar objects in this region, we have 
constructed the colour-colour diagram of J-H vs. H-K$_s$ using the 27 sources 
detected in JHK$_s$ bands. This is shown in Fig. \ref{glimpal} (left). The 
loci of the main-sequence and giant branches are shown by the solid 
and dotted lines, respectively. While the short-dashed line 
represents the locus of 
T Tauri stars \citep{1997AJ....114..288M}, the dot-dashed line 
represents the reddening vector of the main-sequence stars. The long-dashed 
line shows the locus of Herbig Ae/Be stars \citep{1992ApJ...393..278L}. We 
have assumed extinction values of A$_J$/A$_V$=0.282, A$_H$/A$_V$=0.175 and 
A$_K$/A$_V$=0.112 from \citet{1985ApJ...288..618R}. The sources having 
infrared excess are those lying to the right of the reddening curve. We find 
a total of 10 sources having infrared-excess. These sources are included in 
Table \ref{ysolist}. The source associated with IRAS 18511 A has an infrared 
excess as well as very large reddening (J-H$ \sim$4.1, H-K$_s$ $\sim$2.7) as 
can be seen in the CC diagram. IRAS 18511 A is listed as source 
number 7 in Table \ref{ysolist}. 

We have also searched for Spitzer-IRAC counterparts to the near infrared 
objects. We have used a search radius of 0\farcs8. A total of 24 Palomar 
sources have IRAC counterparts. \citet{2007ApJ...659.1360W} have used 
J-[3.6] vs. K$_s$-[4.5] colour-colour diagram to identify the young stellar 
objects in groups around Herbig Ae/Be stars. In our sample we find 6 objects 
that are detected in the J, K$_s$, IRAC1 (3.6~$\mu$m) as well as IRAC2 
(4.5~$\mu$m) bands. These objects have been plotted in the J-[3.6] vs. 
K$_s$-[4.5] colour-colour diagram in Fig. \ref{glimpal} (right). All the 
sources 
in this colour-colour diagram have been dereddened by 7 magnitudes of visual 
extinction (1.8 mag kpc$^{-1}$), corresponding to the extinction due to 
interstellar medium \citep{1992JBAA..102..230W}. We have used 
the \citet{2007ApJ...659.1360W} line (their Eqn. 1) to separate the normal 
stars and the young stellar objects. This is shown by the solid line in the 
Fig. \ref{glimpal}. The dotted line represents the young stellar object (YSO) 
locus given by them. From this 
diagram, we find 3 objects which can be characterised as YSO candidates based 
on the above criterion. The details of the 3 YSO candidates from the 
J-[3.6] vs. K$_s$-[4.5] are included in Table \ref{ysolist}. 
All the young stellar objects selected from various colour-colour diagrams 
([3.6]-[4.5] vs. [5.8]-[8.0], J-H vs. H-K$_s$, J-[3.6] vs. K$_s$-[4.5]) are 
overplotted on the grayscale Palomar K$_s$ band image in Fig. \ref{Pal} and 
listed in Table \ref{ysolist}.

\subsection{Radio emission from ionised gas}

The VLA data show weak emission (point source) at both 8.5 and 15 GHz in 
the IRAS 18511 region. The emission is from a point source with coordinates 
($\alpha_{J2000}$, $\delta_{J2000}$ = ) 18$^h$ 53$^m$ 38\fs67 +01\degr 
50\arcmin 13\farcs0. The location of the radio point source coincides with G5, 
which is marked as a cross in the 8.0~$\mu$m image in Fig. \ref{glim_mips}. The 
flux densities are 0.69 mJy/beam and 0.68 mJy/beam at 8.5 and 15 GHz, 
respectively. The size of the radio emitting region is 
$\la$5\arcsec\ which corresponds to 0.09 pc at the distance of IRAS 18511.
The spectral index is -0.03$^{+0.23}_{-0.30}$ indicating the nature of this 
emission to be (optically thin) free-free emission. Thus, the ionised 
gas here is from a small 
extremely compact region around this source. Using the formulation of 
\citet{1969ApJ...156..269S}, as well as \citet{1973AJ.....78..929P}, 
the ZAMS spectral type of this source is estimated to be B2-B1 (flux 
of Lyman continuum photons is $\sim1.3\times10^{45}$ s$^{-1}$). There 
is no radio emission detected from the other sources in this region, including 
the bright A source, upto $2\sigma$ levels of 100~$\mu$Jy
and 300~$\mu$Jy at 8.5 and 15 GHz, respectively.

\subsection{Source C}
The source C is clearly detected at 24 and 70~$\mu$m and appears as a faint 
diffuse emission. The weak diffuse emission shows filamentary
structures within it. Also, towards the north, we discern a filamentary
structure with a sharp edge. The cause of this emission is unclear.
The emission at 8.0~$\mu$m is stronger than at 24~$\mu$m indicating the presence
of strong Polycyclic Aromatic Hydrocarbons (PAH) emission. The angular size of
C is $\sim$35\arcsec\ which corresponds to a size of 0.7 pc at 3.9 kpc 
(distance to IRAS 18511). While C being an irregular galaxy cannot be ruled 
out, it is possible that C is a Galactic object.

\subsection{Spectral energy distributions}

The fluxes at different wavelengths have been extracted for various sources 
in IRAS 18511 region and these have been used in constructing their  
SEDs. For the sake of comparison, we have also 
constructed the SED of IRAS 18511 as a single source using IRAS as well
as MSX fluxes. The luminosity obtained by integrating the area under the 
IRAS-MSX curve gives an estimate of the total luminosity. 
Figure \ref{seds} (left) shows the SED of IRAS 18511 constructed using IRAS and
MSX fluxes. Also plotted in this figure is the SED of A. It is to be noted
that the fluxes of A at the IRAC and MIPS 24~$\mu$m band are lower limits as A
is saturated in these wavebands. The luminosity of IRAS 18511, obtained by 
integrating the IRAS-MSX SED, is $1.1\times10^4$ \Lsol. The lower limit to
the luminosity of A is $7.3\times10^3$ \Lsol, derived using Spitzer 
photometry and JCMT-SCUBA data. Hence, most of the luminosity of IRAS 18511 
(at least 66\%) is due 
to the protostar(s) in A. 

The SEDs of the individual members (G1-G5) have been constructed using 
wavelengths at which they are resolved and detected.
These SEDs are shown in Fig. \ref{seds} (right) for no 
dereddening applied. The SEDs of these individual objects have been
constructed using the Palomar as well as IRAC fluxes. From Fig. \ref{seds}, we 
see that all the SEDs rise rapidly with increasing wavelength. In order 
to compare 
the SEDs relative to each other, we have normalised them with respect to the 
fluxes in H band. The SEDs of G1, G2, G4 and G5 have been dereddened by 
A$_V$ $\sim$7 magnitudes of visual extinction (due to ISM) as well as by 
A$_V$ $\sim$ 22 mag derived by \citet{2002AJ....124.2790I}. Figure 
\ref{sim_sedAv} shows the SEDs relative to each other for a dereddening of 
A$_V$ $\sim$ 7 mag (left) and 22 mag (right), respectively. By integrating 
the area under the curves (near to mid infrared), we have estimated 
lower-limits to the 
luminosities (L$_{NIR-8\mu m}$) of these objects. These luminosities are
 listed in Table \ref{lum} for the sources G1-G5 for dereddening of 7 and 22 
mag of visual 
extinction. The source G5 appears to be identical to the VLA source 
discussed in Sect. 3.4 and hence we expect it to be of B1-B2 spectral type 
with an effective temperature $\sim20,000$ K and a bolometric
 luminosity 4400 \Lsol.

\subsubsection{Robitaille et al. models}

From Fig. \ref{sim_sedAv}, we see that the sources show a rise in spectral 
energy distributions with an increase in wavelength. Their 
L$_{NIR-8\mu m}$ (listed in Table \ref{lum}) suggest that at 
least a few of these are massive young objects.
In order to get a qualitative estimate of the evolutionary stage of the 
cluster members, we have fitted the SED of few cluster members with the models 
of \citet{2007ApJS..169..328R} (hereafter RWIW). They have computed a 
large set of 
radiation transfer models and obtained the SEDs for a reasonably large 
parameter space. These SEDs can be fitted to multi-wavelength observational 
data of single sources to constrain the physical parameters and the 
evolutionary stage. However, it is important to note that the best constraints 
are obtained if the near, mid and far infrared as well as sub-millimetre data 
are included. 
 
We have fitted the RWIW models to two sources, G4 and G5. The models 
were fit to the SEDs dereddened by 22 mag of visual extinction. The 
L$_{NIR-8\mu m}$s of G4 and G5 are the highest among the G sources. 
The SEDs have been constructed using near infrared Palomar and mid infrared 
IRAC data. Upper limits at MIPS 24~$\mu$m (corresponding to the IRAS 18511 B) 
and at JCMT-SCUBA 850~$\mu$m (flux density at the position of 
IRAS 18511 B) have also been used as constraints to the modelling.
It should be noted that our sub-millimetre angular resolution is not 
sufficient for this purpose and hence the fits are non-unique. We have also 
put additional constraints based on luminosity and mass of envelope/disk 
(85 \Msol\ based on the sub-millimetre 850~$\mu$m flux at the position of IRAS 
18511 B). The models are selected based on the least-squares chi-square method. 
For G5, additional constraints are available since it is estimated to 
be of ZAMS spectral type B1-B2 (based on the VLA observations). Among the four 
models for G5, Model 1 simulates a young protostar and the 
T$_{eff}$ of the central object does not produce Lyman 
continuum photons to ionise the surrounding gas. On the other hand, the other 
models assume more evolved central stars (T$_{eff}$ $\sim$20000 K). 
 Figure \ref{SED_fits} shows some sample RWIW models for G4 and G5 
along with the observed SEDs. 
The parameters are listed in Table 
\ref{rob_parm}. In the Table, column 1 lists the source (G4/G5)
column 2 (M$_*$) lists the mass of central object, column 3
lists the effective temperature of the central objects, column 4 lists the
total luminosity, column 5 lists the inclination angle with respect
to the observer (90\degr is edge-on), column 6 (\.{M}) represents the envelope 
accretion rate, column 7 (M$_{env}$) lists the mass of envelope, column 8 
(M$_{disk}$) lists the mass of disk , column 9 (A$_V$) lists the      
extinction along the line-of-sight and column 10 lists the age. The luminosity 
listed in the Table is the total (non-isotropic) luminosity (central object + 
envelope and/or disk) computed by RWIW. 

It is interesting, especially for G5, that the observed fluxes in the 
mid infrared (IRAC) bands have a steeper wavelength
dependence than predicted by these models which perhaps underestimate the
extinction. Another likely possibility is that emission from small grains or
from Polycyclic Aromatic Hydrocarbons (PAH) effects the observations.
The results indicate that the SEDs of few sources can be explained by massive 
Class I and Class II type sources i.e., sources with the 
massive central objects ($7-10$ \Msol) alongwith envelopes/disks. Thus, the 
models of low-mass objects extended to more massive candidates can reasonably 
explain the SEDs of some cluster members of IRAS 18511.  

\section{Cluster simulations}

In the IRAS 18511 region, a number of sources have been detected in the near 
and mid infrared. In regions of high extinction like IRAS 18511, these 
sources could be highly reddened luminous objects or low-luminosity 
low-extinction objects. The objective of the cluster simulations is to 
investigate this. We simulate a cluster of young objects embedded in a 
cloud of gas and ascertain whether the observations of IRAS 18511 in 
various near and mid infrared bands are in qualitative agreement with the 
predictions of the model as our statistics are too low for a quantitative 
comparison. We explore the cluster membership in terms of the fraction of 
objects of different evolutionary stages (Class I and Class II). As we are 
exploring a young embedded cluster, we have not considered objects in the 
Class III phase or later. This modelling will allow us to obtain a  
qualitative estimate of the evolutionary stage of the cluster by 
varying the fractions of Class I and Class II sources in the cluster.

The inputs to these simulations include (a) observables - quantities that have 
been incorporated in the model based on observations, and (b) assumptions.  
The observables include (1)  limit on the most massive object in the cluster, 
(2) the total luminosity of the cluster, (3) mass and size of the spherical 
cloud of
gas in which the cluster is embedded, and (4) the distance. We have assumed 
the following: (1) Initial Mass Function (IMF) and (2) Star formation history  
(SFH). The output that can be compared directly with the 
observations is the number of detected cluster members in the K$_s$ 
(2.12~$\mu$m), IRAC1 (3.6~$\mu$m), IRAC2 (4.5~$\mu$m), IRAC3 (5.8~$\mu$m), 
IRAC4 (8.0~$\mu$m) and MIPS1 (24~$\mu$m) bands for given completeness limits.
A detailed description of the model is given in Appendix B.

\subsection{Results}

\subsubsection{Model}

The upper mass limit of the cluster has been taken to be 10 \Msol\ 
corresponding to the ZAMS spectral type of a \textit{single} object 
inferred from the bolometric luminosity of 
IRAS 18511. The model has been run for the Salpeter IMF. The mass and radius of 
the spherical cloud of gas have been taken to be 750 \Msol\ and 0.6 pc, 
respectively. These have been derived from the 850~$\mu$m JCMT-SCUBA 
observations. This corresponds to a maximum visual extinction of 69 magnitudes 
(including the $A_V\sim7$ magnitudes due to the ISM).
The total luminosity has been taken to be $1.1\times10^4$ \Lsol\ which is the 
total luminosity of IRAS 18511. The model was run 1000 times for 
each of the following cases of star formation history: 

\begin{enumerate}
\item Coeval formation with an age of 0.5 Myr corresponding to Class I type sources - Every cluster member is assumed to have a Class I type spectrum.
\item Coeval formation with an age of 1 Myr corresponding to Class II type 
sources - Every cluster member is assumed to have a Class II type spectrum.
\item Uniform star formation rate between $0.01 - 1$ Myr - This corresponds to 
the general case incorporating Class I as well as Class II objects. For 
every cluster member of a certain mass, an age is randomly assigned. If this is less than 0.5 Myr, its spectrum is taken to be of the Class I type. For 
ages greater than 0.5 Myr, the cluster member is a Class II type object.
\end{enumerate}

The median number of cluster members is $\sim300$ while the median mass of the 
cluster is $\sim230$ \Msol. We have also estimated the magnitude distribution 
from the cluster for the case of a 
uniform star formation rate (Case 3). This is shown for K$_s$ band apparent 
fluxes in Fig. \ref{fluxd} for the general case incorporating Class I 
and Class II objects with ages between 0.01 and 1 Myr.

\subsubsection{Comparison with observations}

A comparison of the model results with the observations has been carried out in
 terms of magnitude distribution (Fig. \ref{fluxd}) and number of sources 
detected  (Fig. \ref{comp}). Figure \ref{fluxd} shows the simulated (Case 3) 
as well as observed magnitude distributions from the Palomar NIR data. The 
observed sources used in the figure are the young stellar objects detected in 
the K$_s$ band and listed in Table \ref{ysolist}. 
Also marked on the figure is the Palomar K$_s$ band 
sensitivity limit. A comparison of the observed and simulated magnitudes  
 shows that the sources detected in observations are brighter than 
those simulated. This could be because the density distribution of the cloud 
is inhomogeneous while the cluster simulations assume a constant density 
distribution of gas. Another possibility is that the distribution of 
young stellar objects is more extended than the size of the cloud. 
From the simulations, it is clear that with the 
sensitivity limit of Palomar, only a small fraction of sources in the 
cluster are detected. 

From the model, the median number of detected sources in each band has been 
derived based on the completeness limit of the instrument used for 
comparing the results. Figure \ref{comp} shows the number of detected cluster 
members predicted by the model in each band. The errorbars indicate 
the quartile 
values encompassing 50\% of the number of detected cluster members. The 
cross and dotted line represent Case 1 (Class I); open circle 
and dashed line represent Case 2 (Class II); and, filled square and 
dot-dashed line represent Case 3 (general case incorporating Class I
 Class II objects). The observed number of sources (from Palomar image and 
IRAC1, IRAC2, IRAC3 \& IRAC4 bands of GLIMPSE images as well as from MIPS 
24~$\mu$m of MIPSGAL image) are shown by the solid circles 
and solid lines. The 
wavelengths for different cases have been slightly shifted for better viewing. 

The observed number of sources in each band is shown by a solid line depicting 
a range of possible number of cluster members, between 
the lower and upper limits. The lower limit is obtained from the number of 
young stellar objects detected in that band within the `region of interest'
 by using the colour-colour diagrams (ref Table \ref{ysolist}). The 
upper edge of each solid line represents the total number of sources 
detected within that band. For the case of the MIPS 24~$\mu$m band, 
there are only two sources detected within the region of 
interest. IRAS 18511 B shows an extension at 24~$\mu$m and therefore we 
have plotted the number of sources detected as 2. 
 Obviously, this represents a lower limit as the sources are 
unresolved due to lower angular resolution. 

From Fig. \ref{comp}, we find that among all the three cases, the general case 
incorporating Class I ($\sim$50\%) and Class II ($\sim$50\%) sources agrees 
well with the observations. The models with Class I and Class II sources 
alone are inconsistent with the observational data. We, therefore 
infer that the sources in IRAS 18511 are very young and deeply embedded. 
In K$_s$ as well as in IRAC1 bands, the general case (Case 3) as well as 
Class II (Case 2) model results are consistent with observations. 
This is expected as more Class II sources should be 
detected in the K$_s$ band. However, in the IRAC2 and 
IRAC3  bands, all the model results agree with observations within the quartile
 values. For the IRAC4 band, the Class I as well as the general case 
agree with observations better than with the Class II case. Finally, 
at 24~$\mu$m, the number of detected sources from the Class I model 
and the general case are higher 
than that detected from the Class II model. It must be noted that the 
observations at 24~$\mu$m represent a lower limit due to the low angular 
resolution of the MIPS instrument ($\sim$6\arcsec). Among all the cases, the 
general case (Case 3) assuming a mix of $\sim$50\% Class I and $\sim$50\% 
Class II sources can be said to be a reasonable fit to observations. 
The model suffers from a few limitations which are listed below.
\begin{itemize}
\item{The cluster is embedded in
a cloud of gas which is homogeneous and clumpiness/density distributions has
not been taken into consideration.}
\item{The number of detected cluster members from the simulations (which are 
point-like) is compared with the observed number of sources. A better 
comparison would be possible if the results of the simulations were convolved 
with the beam size of the instrument.}
\end{itemize}

\section{Discussion}

\subsection{Clustering}

We have attempted to study the clustering of the sources around IRAS 18511  
using the NIR Palomar images. The method used is described in detail by 
\citet{1998A&AS..133...81T} to study the clustering around Herbig Ae/Be stars. 
A density profile is computed centred on the bright IRAS 18511 A source to 
look for evidence of clustering. Density profiles are 
computed using the Palomar K$_s$, IRAC2 (4.5~$\mu$m) and IRAC4 (8~$\mu$m) band 
sources as shown in Fig. \ref{radprof}. In this plot, 
the star densities derived for the IRAC2 and IRAC4 bands have been 
scaled by factors of 1.7 and 6.4, respectively for comparison with the radial 
profile derived for the K$_s$ band. Further, the radii of the annuli of the 
IRAC bands are slightly shifted for better viewing of the plot.  
From the profile, we do not see any 
`clear' evidence for clustering centred on A.
This is consistent with the results of the previous sections, i.e. the 
extinction due to the associated molecular cloud is rather high and there 
may be other embedded sources. This is consistent with 
the results of our simulations, i.e. only a small fraction of sources 
are detected.

\subsection{Extinction at 8.0~$\mu$m vs. 850~$\mu$m}

A comparison of the morphology of the sub-millimetre dust emission
and the filamentary structures seen in the 8.0~$\mu$m emission map 
indicates that these filamentary structures comprise of cold dust. 
Figure \ref{IRD} shows the 8.0~$\mu$m grayscale image with 850~$\mu$m contours 
overlaid. The morphology of the 850~$\mu$m
emission traces the filamentary structures close to IRAS 18511. It is 
evident that these filaments represent the cold dust comprising the
infrared dark cloud with IRAS 18511 located at the peak of one clump.

We have compared the extinction values obtained from the JCMT-SCUBA 850~$\mu$m 
image (using Eqn. 2) with that from IRAC 8.0~$\mu$m image (using Eqn. 3).
Resampling the images and a pixel-to-pixel comparison of the extinction values
in the region of overlap of the filament (to the south-west of IRAS 18511 A) 
shows that the visual extinction values obtained using the JCMT-SCUBA 
sub-millimetre image is $\sim8$ times the the A$_V$ values obtained from the 
IRAC 8.0~$\mu$m image. However the A$_V$ values quoted for 
the IRAC 8.0~$\mu$m image are  lower limits for the following reason.
Equation 3 assumes the emitting dust to be behind the absorbing filaments
and this is clearly not correct since $I_o$ has contribution
from foreground emission. Consequently, the values of A$_V$ derived are an 
underestimate. And also, the mass estimates derived 
for the filaments and the clump in Sect. 3.2.1 are lower-limits. 
\citet{2001A&A...370..230B} have imaged a number of 
molecular lines in the IRAS 18511 region ($^{13}$CO, CS, HCO$^+$). Of 
particular 
interest is the HCO$^+$ emission integrated over the red part of the emission.
This emission extends along the direction of the filaments to the south-west, 
as seen in the JCMT-SCUBA 850~$\mu$m image. Further, the emission is at a
velocity (V$_{lsr}$) of $\sim59$ km s$^{-1}$. This is the  V$_{lsr}$ of IRAS 
18511 and hence, the filamentary structures are associated with the molecular 
cloud of IRAS 18511.    

Using the fact that the A$_V$ values derived from the 850~$\mu$m 
sub-millimetre map is 
$\sim$8 times larger than that obtained from the IRAC 8.0~$\mu$m map, we can 
estimate the fraction of foreground emission contributing to the total 
emission detected (in the neighbourhood of the filaments). We estimate that 
this fraction of foreground emission is $\sim$80\% of the detected emission. 
In other words, most of the emission is foreground emission. This is not 
surprising considering that IRAS 18511 is a distant object ($\sim$3.9 kpc).

\subsection{Evolutionary stage of cluster members}
It is interesting to compare the multiwavelength observations of A and B,
although each comprises of more than one source. A is the brightest
source in the near, mid and far infrared as well as in the sub-millimetre
wavebands. However, there is no detected radio emission associated with A. On 
the other hand, G5 (associated with B) has free-free radio emission 
associated with it. The peak of molecular line (C$^{18}$O and $^{13}$CO) 
emission, as observed by \citet{1999ApJS..125..143W} is located
near B and one can see an extension corresponding to A. The absence of radio
emission near A and other observations suggest the following possibilities:
(1) A is in an earlier evolutionary stage than B and the ultracompact 
\ion{H}{ii}
region has not yet formed. This would explain the absence of radio emission
as well as their brightness at infrared wavebands (2) A being more 
massive than B, its
ultracompact \ion{H}{ii} region is being choked by the infalling matter through
accretion \citep{2003ApJ...599.1196K}. \citet{2004ApJS..155..149K} have 
carried out a VLA survey in the 
44 GHz (Class I) methanol maser line in several star forming regions 
including IRAS 18511. They find a maser in the vicinity of IRAS 18511 A 
(synthesized beam size $\sim2\farcs2\times1\farcs5$). The position of the 
maser (positional uncertainy $\sim0\farcs5$) is shown in Fig. 
\ref{glim_mips}. They also find in their survey that Class I methanol masers 
are typically offset by $\sim0.2$ pc (median value) from other 
massive star formation signposts like \ion{H}{ii} regions or water masers. 
\citet{2007arXiv0704.0988C} have studied the massive protocluster 
associated with S255N and they find that the Class I methanol are  
aligned in the direction of the outflow.
 In the case of IRAS 18511, the methanol maser is at distance of 0.1 pc from 
IRAS 18511 A and 0.4 pc from G5 (VLA source). Also, no water masers or 
\ion{H}{ii} 
region have been detected near IRAS 18511 A. Further, it is believed 
that shocked gas from outflows/accretion might be responsible for 
Class I methanol masers 
\citep{1990ApJ...364..555P,1992ApJ...385..232J}. \citet{2005ApJ...625..864Z} have imaged IRAS 18511 ($\sim$30\arcsec) in the CO (J=$2\rightarrow1$) line and 
the emission shows a peak close to IRAS 18511 A. 
This is unlike the case of C$^{18}$O where the peak of emission is located at 
IRAS 18511 B \citep{1999ApJS..125..143W}. There is also evidence for an 
outflow from IRAS 18511 A since wings are seen in the line profiles observed 
by \citet{2001A&A...370..230B}. All these facts suggest that IRAS 
18511 A is in an earlier evolutionary state than IRAS 18511 B.

Considering the above observations, analyses and simulations of the cluster 
associated with IRAS 18511, the following likely scenario emerges. This is a 
young cluster with sources in an early evolutionary phase (Class I 
and Class II). The large extinction indicates that probably there are more 
low-mass cluster members hidden in the molecular cloud which will start 
appearing as the cluster evolves and the molecular cloud disperses. IRAS 18511 
A is likely to be an intermediate mass/massive young object, probably 
a Class I object. Its luminosity indicates that it is probably the most 
massive object in the cluster. On the other hand, the presence of radio 
emission near G5 indicates that G5 could be an older member of the 
cluster. The absence of other young 
objects in the vicinity of IRAS 18511 A can possibly be explained by large 
extinction as indicated by the JCMT sub-millimetre maps (peak of emission is at 
A). Further study of IRAS 18511 (high angular resolution mid infrared 
maps and molecular line maps) can shed light on the likely scenario. 

\subsection{Precursor to Herbig star type cluster?}

The clusters associated with Herbig stars and their properties have been 
studied by \citet{1997A&A...320..159T}, \citet{1998A&AS..133...81T}, 
\citet{1999A&A...342..515T}. In particular, the mid infrared emission 
(10~$\mu$m) from a few such clusters has been 
investigated by \citet{2003A&A...400..575H}. The luminosity as well as the 
clustering suggests that IRAS 18511 could be a precursor to a Herbig cluster. 
\citet{1999A&A...342..515T} report the tendency of early Herbig Be stars to be 
surrounded  by dense clusters of lower mass companions. Some of the clusters 
associated with earlier Herbig stars (of ZAMS spectral type B0) are found to 
have $\sim75$ members, where the lower mass limit is $\sim0.2$ \Msol\ for 2 
magnitudes of extinction in K. It is interesting to compare these numbers 
with the model results which suggest cluster membership to be $\sim200$ for
 masses greater than 0.2 \Msol. While the model results predict the number of 
cluster members to be $\sim2-3$ times the number observed by 
\citet{1999A&A...342..515T}, it is useful to note that the clusters of 
\citet{1999A&A...342..515T} are more evolved as compared to the model which is 
an embedded cluster. And evolved clusters with intermediate mass Herbig Ae/Be stars retain less than 50\% of its members \citep{2007MNRAS.376.1879W} 
Further, the size of the embedded cluster 
(size of cloud from sub-millimetre map) associated with IRAS 18511 is $\sim$0.6 
pc. This compares well with the cluster sizes of 0.2-0.7 pc obtained for 
clusters associated with Herbig Ae/Be stars of spectral type B0 
\citep{1999A&A...342..515T}.
The mid infrared (excess) emission as well as the large extinction (it is not 
visible optically) indicates that the main source (A) is in an early 
evolutionary stage. \citet{2003A&A...400..575H} find objects with mid infrared 
emission 
from circumstellar  disks and envelopes in five out of twelve fields of known 
Herbig AeBe stars they studied. The IRAS 18511 field also shows many objects in 
mid infrared. In short, the following considerations: (a) luminosity 
(b) evolutionary stage of the objects (c) size of associated cluster and (d) 
number of cluster members, suggest that the source associated with 
IRAS 18511 is a protocluster associated with a candidate precursor to 
a Herbig-star cluster. This is in 
accordance with the suggestions of \citet{1998A&AS..133...81T} and 
\citet{2000A&A...355..617M}. We therefore, believe that IRAS 18511  
 represents an early stage of clusters associated with an 
intermediate mass object.

\section{Summary}

With the aim of studying the early evolution of clusters, we have selected 
IRAS 18511 to carry out a detailed investigation. Using emission at 
sub-millimetre (JCMT-SCUBA), infrared (Spitzer-MIPS, Spitzer-IRAC 
and Palomar) and radio (VLA) wavelengths, we have studied the main source as 
well as the cluster associated with IRAS 18511. The results are consistent 
with simulations of a young embedded cluster incorporating Class I and Class 
II sources. Based on the luminosity, and properties of the associated cluster 
(number of members, size of cluster and evolutionary stages of objects), we 
conclude that IRAS 18511 is a protocluster with the most massive object
being a candidate precursor to a Herbig Ae/Be star.

\begin{acknowledgements}
We would like to thank R. Cesaroni for his help and suggestions.
\end{acknowledgements}

\bibliography{ref}

\begin{table*}
\caption{The MIPS flux density details of sources in IRAS 18511.}
\label{fluxes}
\begin{tabular}{|c | c| c | c|} \hline \hline
 Source & Position & \multicolumn{2}{c|}{Flux Density (Jy)} \\ \hline
 & J2000 & 24~$\mu$m & 70~$\mu$m \\ \hline
A &  18$^h$ 53$^m$ 37\fs85 +01$^d$ 50$'$ 30\farcs5 & $>20.6$ & 145.7 \\
B &  18$^h$ 53$^m$ 38\fs44 +01$^d$ 50$'$ 15\farcs1 & 6.4 & 61.2 \\
C &  18$^h$ 53$^m$ 43\fs55 +01$^d$ 49$'$ 50\farcs4 & 0.8 & 22 \\ \hline
\end{tabular}
\end{table*}

\begin{landscape}
\vskip -5cm
\begin{table}
\caption{Young stellar objects around IRAS 18511 selected from Palomar (NIR) as well as Spitzer-IRAC (MIR) colour-colour diagrams.}
\label{ysolist}
\begin{tabular}{l c c c c c c c c c c} \hline \hline
S. No. & $\alpha_{2000}$ & $\delta_{2000}$ & J & H & K$_s$ & Spitzer-GLIMPSE & 3.6~$\mu$m & 4.5~$\mu$m & 5.8~$\mu$m & 8.0~$\mu$m \\
       & (deg)  & (deg) & (mag) & (mag) & (mag) & designation & (mag) & (mag) & (mag) & (mag) \\ \hline
1 & 283.400479 & 1.841550 & $18.36\pm0.20$ & $17.22\pm0.13$ & $16.50\pm0.19$ & - & - & - & - & - \\
2 & 283.403321 & 1.840394 & $18.86\pm0.36$ & $17.58\pm0.21$ & $15.73\pm0.14$ & - & - & - & - & - \\
3 (G1) & 283.405912 & 1.834686 & - & $16.45\pm0.09$ & $13.43\pm0.02$ & G034.8140+00.3504 & $10.91\pm0.11$ & $10.24\pm0.11$ & $9.80\pm0.32$ & $9.30\pm0.10$ \\
4 & 283.405946 & 1.840603 & $11.85\pm0.01$ & $11.58\pm0.01$ & $11.11\pm0.01$ & - & - & - & - & - \\
5 & 283.406588 & 1.840244 & $13.12\pm0.01$ & $12.67\pm0.01$ & $11.89\pm0.01$ & - & - & - & - & - \\
6 (G2) & 283.407446 & 1.833108 & $18.64\pm0.37$ & $15.89\pm0.08$ & $14.88\pm0.07$ & G034.8133+00.3483 & $13.19\pm0.11$ & $12.57\pm0.14$ & $11.18\pm0.24$ & $9.57\pm0.10$ \\
7$^a$ & 283.407867 & 1.841829 & $13.43\pm0.04$ & $9.30\pm0.03$ & $6.61\pm0.03$ & - & - & - & - & - \\
8 (G3) & 283.409387 & 1.835869 & - & - & $14.67\pm0.08$ & G034.8166+00.3479 & $11.00\pm0.21$ & $10.41\pm0.21$ & $7.84\pm0.16$ & $6.09\pm0.27$ \\
9 & 283.409779 & 1.842842 & $15.95\pm0.04$ & $15.00\pm0.03$ & $13.38\pm0.02$ & - & - & - & - & - \\
10 (G4) & 283.409896 & 1.837536 & - & $15.00\pm0.03$ & $11.11\pm0.01$ & G034.8183+00.3482 & $7.56\pm0.08$ & $6.58\pm0.19$ & $5.65\pm0.05$ & $4.95\pm0.07$ \\
11 (G5) & 283.411304 & 1.837044 & - & $17.54\pm0.23$ & $12.77\pm0.02$ & G034.8185+00.3467 & $7.61\pm0.05$ & $6.28\pm0.20$ & $5.11\pm0.04$ &  $4.32\pm0.04$ \\
12 & 283.411446 & 1.832517 & $17.69\pm0.15$ & $17.21\pm0.21$ & $16.64\pm0.26$ & - & - & - & - & - \\
13 & 283.412083 & 1.833464 & $17.19\pm0.09$ & $15.55\pm0.04$ & $14.16\pm0.03$ & - & - & - & - & - \\
14 & 283.414129 & 1.839558 & $17.88\pm0.14$ & $17.31\pm0.14$ & $16.86\pm0.32$ & - & - & - & - & - \\
15 & 283.415887 & 1.835761 & $14.53\pm0.02$ & $13.93\pm0.02$ & $13.66\pm0.02$ & G034.8194+00.3420 & $13.42\pm0.17$ & $13.20\pm0.13$ & - & - \\
16 & 283.415892 & 1.834844 & $18.02\pm0.16$ & $16.28\pm0.06$ & $14.66\pm0.04$ & G034.8187+00.3417 & $13.15\pm0.14$ & $12.81\pm0.12$ & - & - \\
\hline
\end{tabular}

$^a$ Source saturated in the GLIMPSE-IRAC images \\
\end{table}
\end{landscape}

\begin{table*}
\caption{Luminosities of the sources G1-G5 in IRAS 18511 region detected in all 
four IRAC bands}
\label{lum}
\begin{tabular}{c c c } \hline \hline
S. No. & \multicolumn{2}{c}{Luminosity (L$_{NIR-8\mu m}$) } \\ 
&  \Lsol\ (A$_V\sim7$ mag) & \Lsol\ ($A_V\sim22$ mag) \\ \hline
G1 &  7   &  16 \\
G2 &  2   &   9 \\
G3 &  19  &  36 \\
G4 & 184  & 412 \\
G5 &  238 & 492 \\
\hline
\end{tabular}
\end{table*}

\begin{table*}
\caption{Parameters of the \citet{2007ApJS..169..328R} models for G4 
and G5, plotted in Fig. 
\ref{SED_fits}. Column 2 (M$_*$) lists the mass of central object, column 3 
lists the effective temperature of the central objects, column 4 lists the 
total luminosity, column 5 lists the inclination angle with respect 
to the observer (90\degr is edge-on), column 6 (\.{M}) represents the envelope 
accretion rate, column 7 (M$_{env}$) lists the mass of envelope, column 8 
(M$_{disk}$) lists the mass of disk , column 9 (A$_V$) lists the 
envelope extinction along the line-of-sight and column 10 lists the 
age.}
\label{rob_parm}
\begin{tabular}{c c c c c c c c c c} \hline \hline
Obj. & M$_*$ & T$_{eff}$ & Luminosity & Incl. angle & \.{M} & M$_{env}$ & 
M$_{disk}$ & A$_V$ & Age \\ 
 & (\Msol) & (K) & (\Lsol) & (deg) & (\Msol/yr) & (\Msol) & (\Msol) & (mag) & (yr) \\ \hline

G4 Model 1& 6.7 & 4241 & $5.2\times10^2$ & 18 & $3.9\times10^{-5}$ & 2.4 & $1.6\times10^{-2}$ & 19.3 & $5\times10^3$ \\
G4 Model 2& 9.2 & 24552 & $4.6\times10^3$ & 81 & 0 & - & $2.1\times10^{-2}$ & 24 & $1\times10^6$ \\
G5 Model 1 & 10.1 & 4351 & $2.2\times10^3$ & 18 & $1.1\times10^{-3}$ & 47.5 & $4.0\times10^{-2}$ &1960 & $4\times10^3$ \\
G5 Model 2 & 7.5 & 21770 & $2.3\times10^3$ & 87 & $1.1\times10^{-6}$ & 1.3 & $8.4\times10^{-3}$ & 414 & $4\times10^5$ \\
G5 Model 3 & 7.3 & 21262 & $2.0\times10^3$ & 87 & $8.6\times10^{-9}$ & 0.07 & $2.9\times10^{-5}$ & 12 & $1\times10^6$ \\
G5 Model 4 & 8.1 & 22602 & $2.9\times10^3$ & 87 & 0 & - & $1.6\times10^{-3}$ & $5.5\times10^{-4}$ & $4\times10^6$ \\
\hline
\end{tabular}

The cavity angle in Model 1 for both G4 and G5 is 
$2^{\circ}-6^{\circ}$ while the cavity angle in Models 2 and 3 of G5 is
$40^{\circ}-50^{\circ}$. \\
\end{table*}

\clearpage

\begin {figure*}
\includegraphics[height=9.0cm]{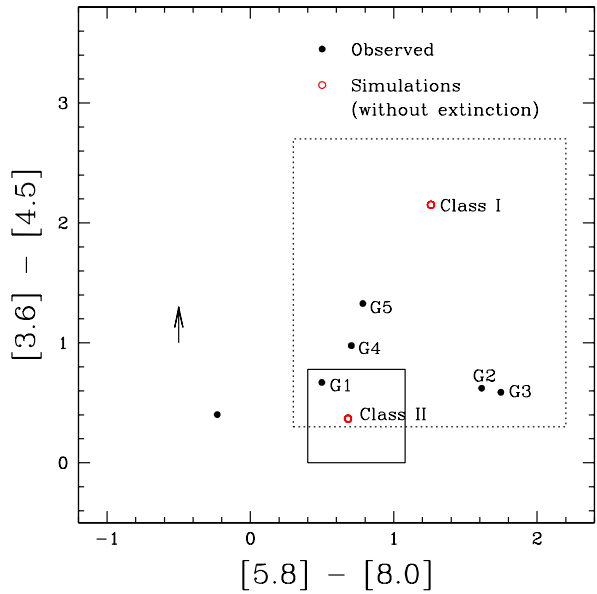}
\caption{Spitzer colour-colour diagram of the observed sources and the models 
used in the simulations. The filled circles represent the sources detected 
in all four IRAC bands. The empty (red) circles represent the colours of the 
unreddened Class I and Class II models used in the simulations (see text for 
details). The solid line square approximately delineates the
region occupied by Class II sources whereas the dotted-line square covers the
region occupied by the Class I models shown in the models of Allen et al. 
(2004). The arrow represents the extinction vector for A$_V$=20 magnitudes.
}
\label{ccd}
\end {figure*}

\begin {figure*}
\vskip -3.5cm
\includegraphics[height=15.0cm]{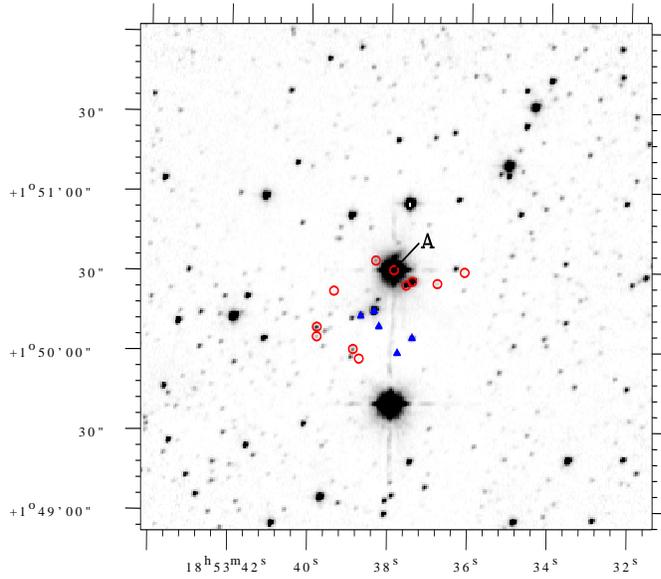}
\vskip -3.5cm
\caption{K$_s$ band Palomar image of the region around IRAS 18511. The axes
are in J2000 coordinates. The sources marked by open (red) circles are the
YSO candidates selected from the various colour-colour diagrams (see Table
\ref{ysolist}). The (blue) solid triangles are the sources G1-G5.
}
\label{Pal}
\end {figure*}

\begin {figure*}
\includegraphics[height=9.0cm]{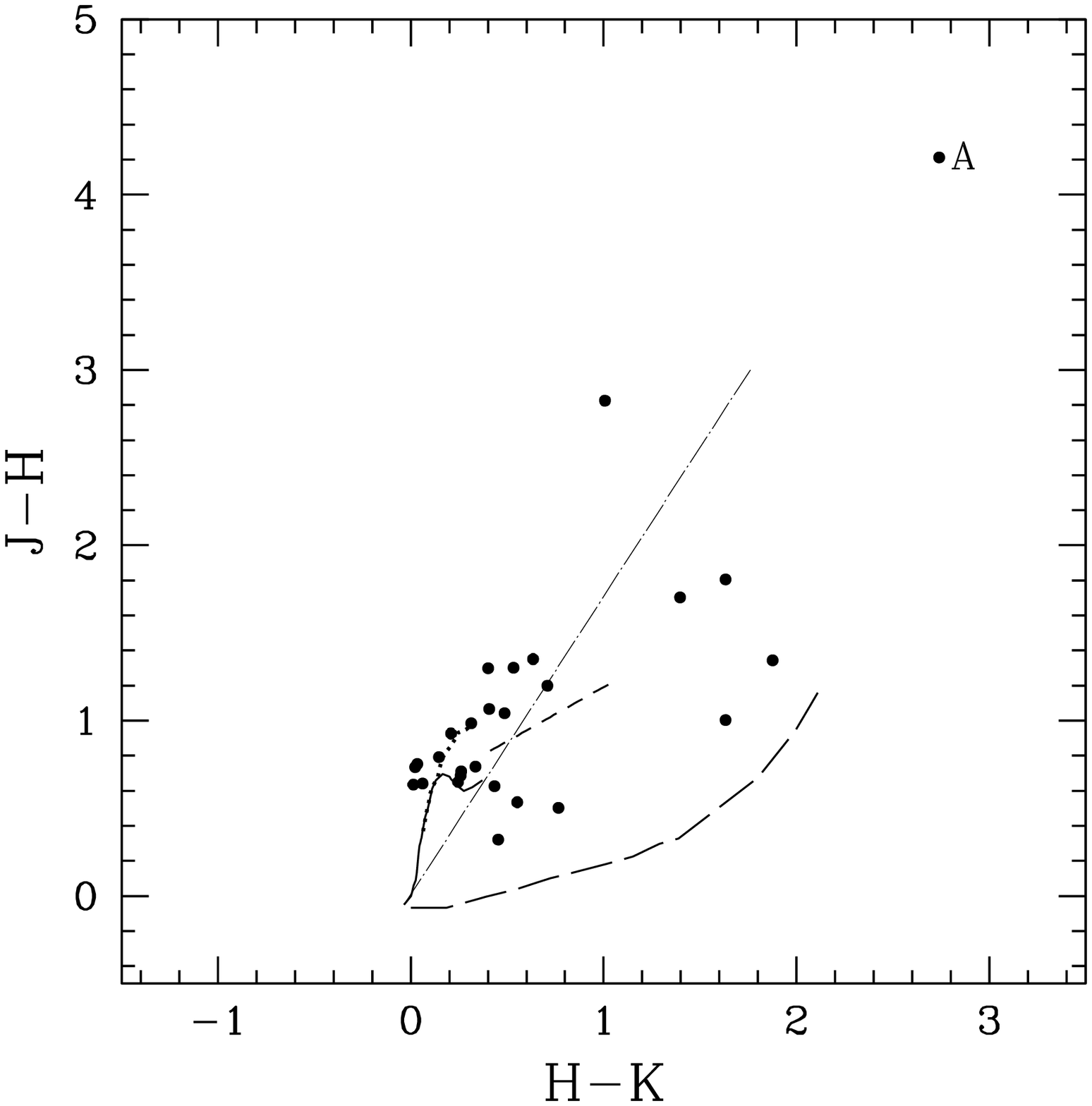}
\includegraphics[height=9.0cm]{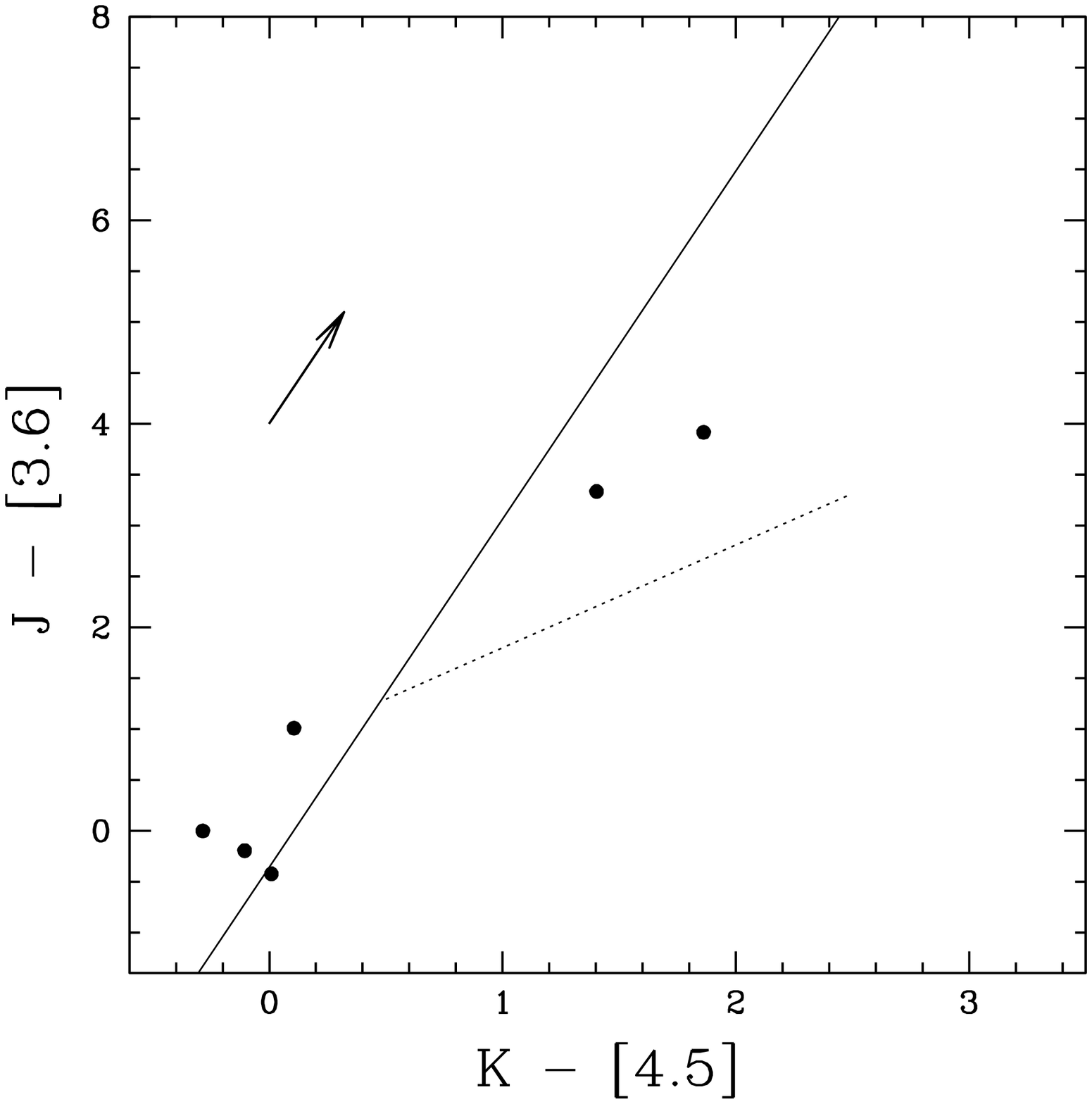}
\caption { Colour-colour diagrams (CCDs) of sources detected in Palomar and 
Spitzer-IRAC images within 10\% contour level of the sub-millimetre 850~$\mu$m 
peak at IRAS 18511 A (see text for details). 
(Left) J-H vs. H-K$_s$ CCD where the locii of the
main sequence and giant branches are shown by the solid and dotted lines
respectively. The short-dash line represents the locus of T-Tauri stars. The
 dash-dotted straight line follows the reddening vectors of 
main sequence stars (or dwarfs). The long dashed line represents the locus of 
Herbig Ae/Be stars. The position of IRAS 18511 A is marked as `A' in the CCD. 
(Right) J-[3.6] vs. K$_s$-[4.5] CCD where the solid line represents the line 
separating the normal stars and the young stellar objects while
the dotted line represents the YSO locus of \citet{2007ApJ...659.1360W}. The 
arrow represents the extinction vector A$_V$=5 magnitudes.
}
\label{glimpal}
\end {figure*}

\begin {figure*}
\includegraphics[height=9.0cm]{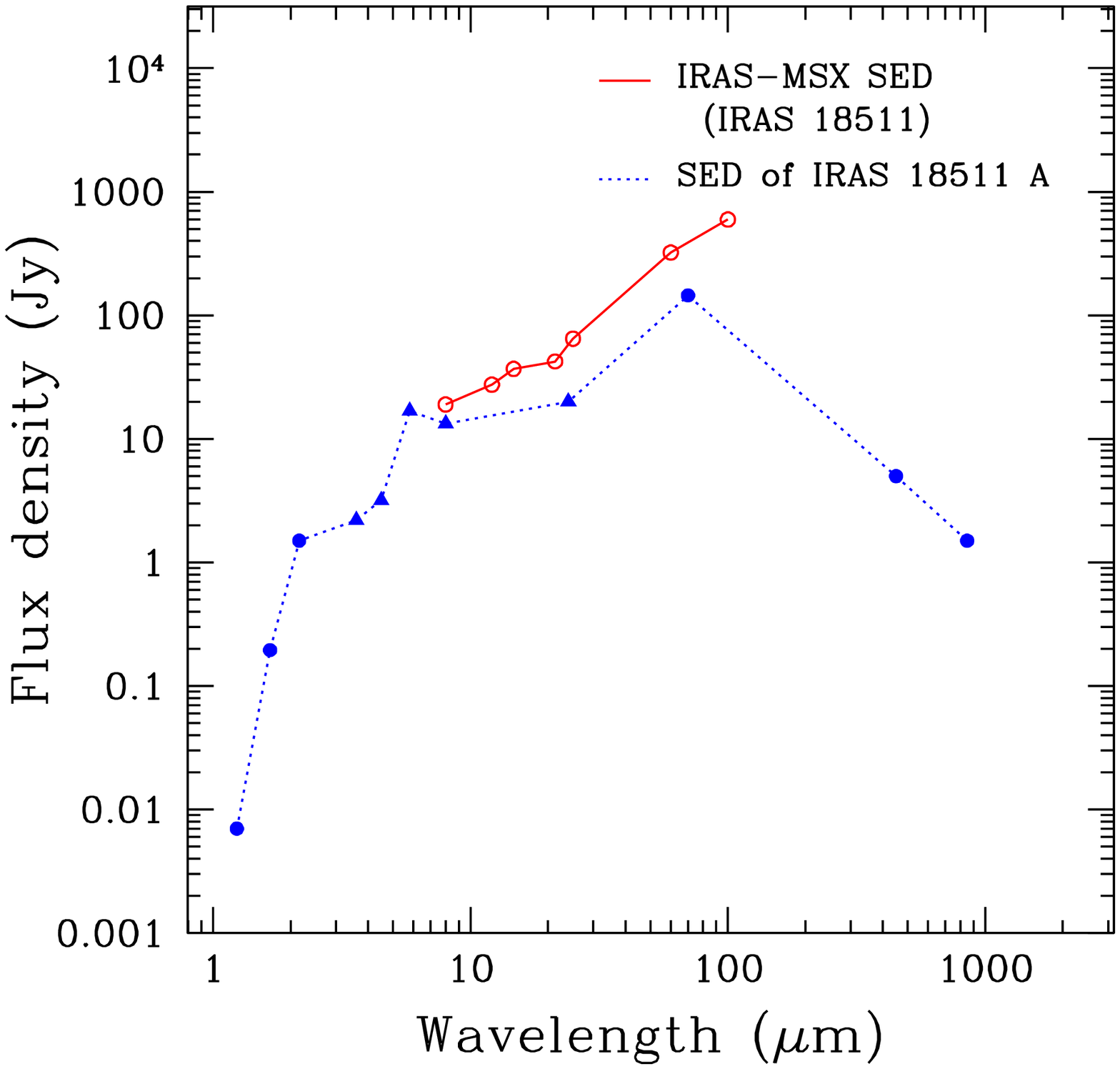}
\includegraphics[height=9.0cm]{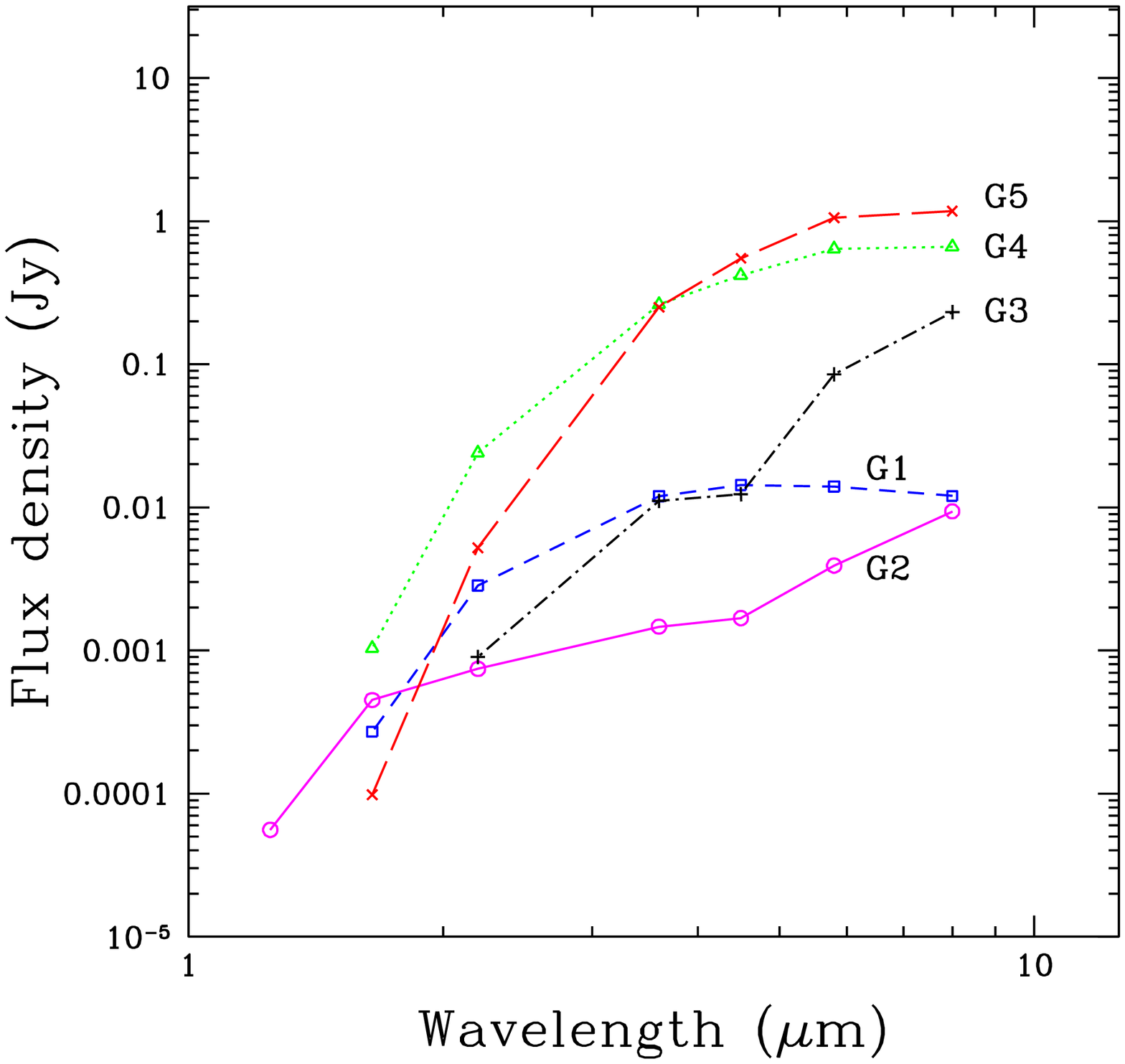}
\caption{(Left) The solid line represents the spectral energy 
distribution (SED) of IRAS 18511 constructed using the IRAS and MSX fluxes. 
The dotted line depicts the lower limit to the spectral energy 
distribution of A constructed using Palomar, Spitzer-IRAC, Spitzer-MIPS and 
JCMT-SCUBA.  While the filled circles represent the fluxes of IRAS 18511 A, the
 solid triangles represent the lower limits. (Right) The SEDs of 
young stellar objects, G1-G5, detected using IRAC colour-colour diagram.
}
\label{seds}
\end {figure*}

\begin {figure*}
\includegraphics[height=9.0cm]{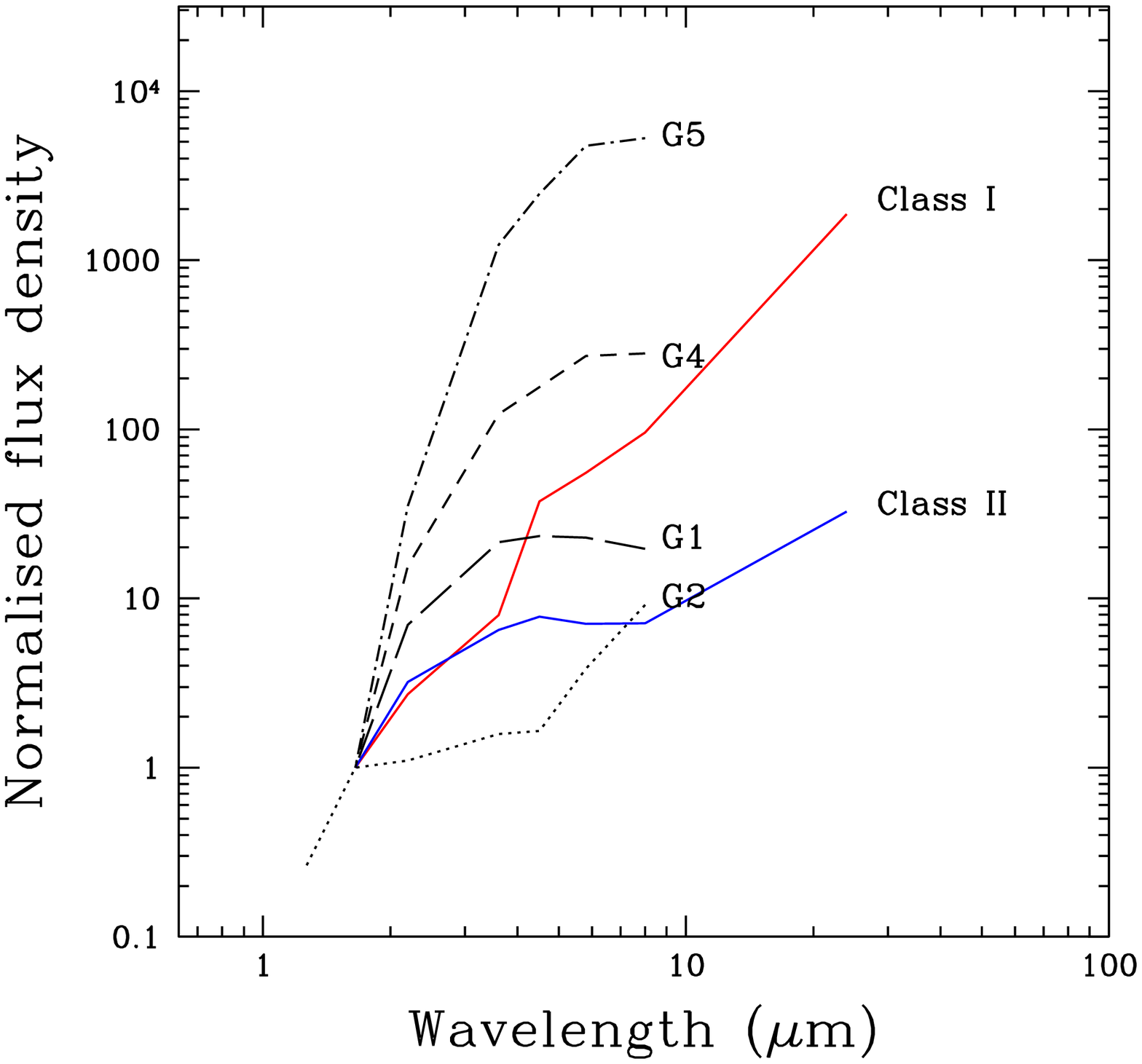}
\includegraphics[height=9.0cm]{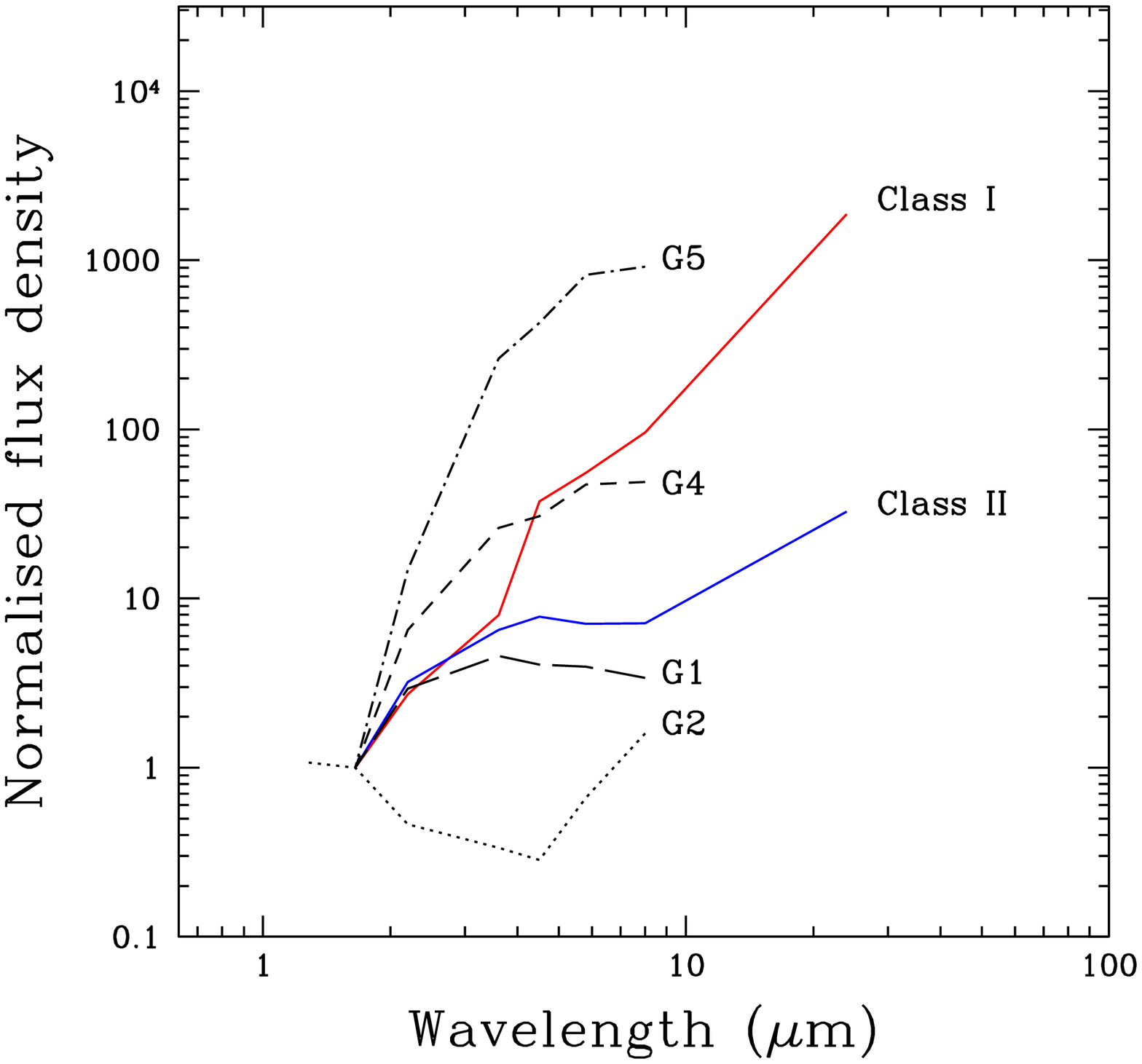}
\caption{Normalised spectral energy distribution (with respect to H band) of 
sources G1, G2, G4 and G5, dereddened by A$_V$ $\sim$7 magnitudes (left) and 
A$_V$ $\sim$22 magnitudes (right). The solid lines represent the normalised 
spectral energy distributions of the Class I (red) and Class II (blue) sources 
adopted in the cluster simulations.
}
\label{sim_sedAv}
\end {figure*}

\begin {figure*}
\includegraphics[height=9.0cm]{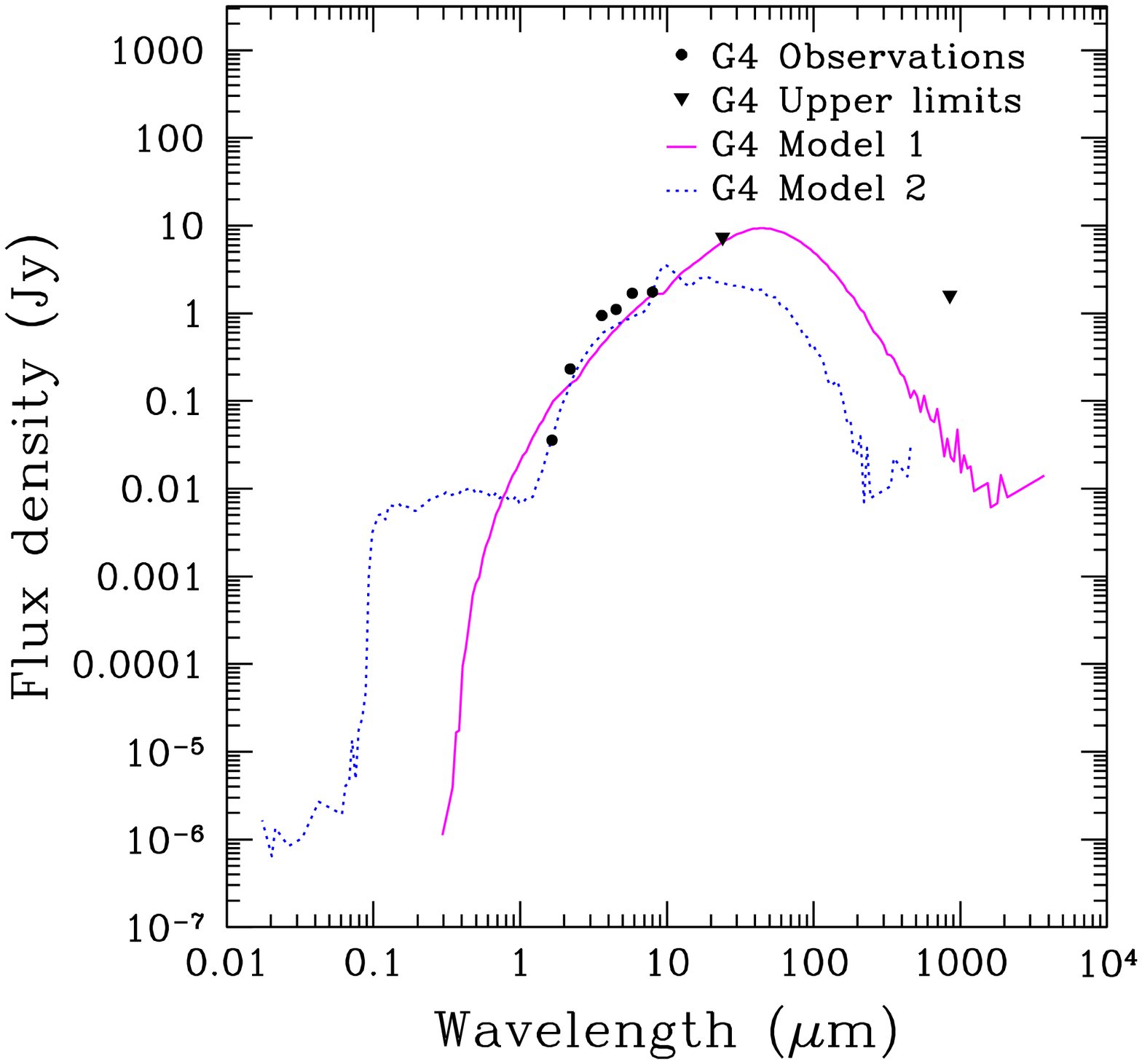}
\includegraphics[height=9.0cm]{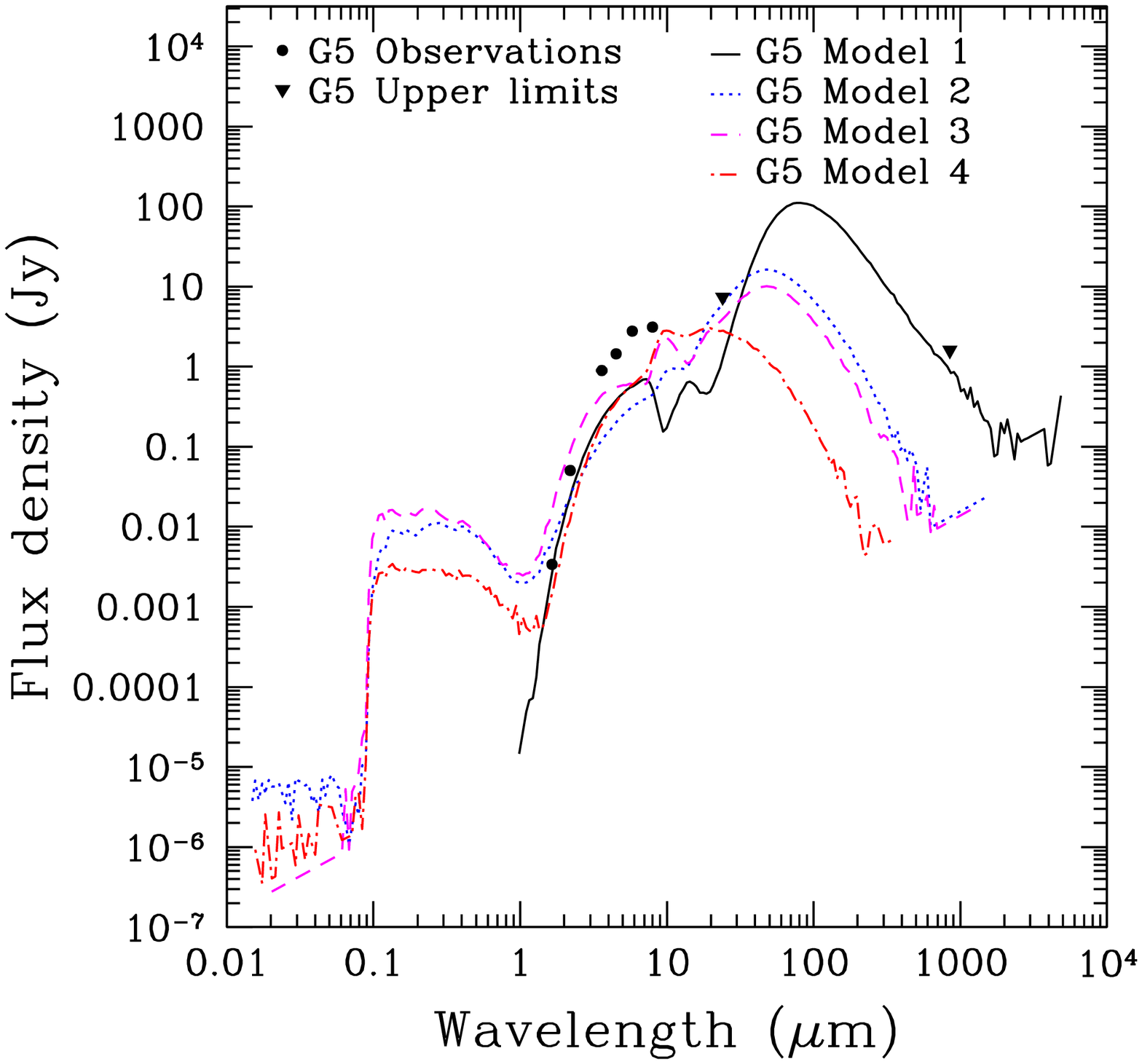}
\caption{Spectral energy distributions of G4 (left) and G5 (right) with 
some sample models of \citet{2007ApJS..169..328R} (RWIW). The 
observed fluxes are shown by filled circles, the triangles denote upper 
limits to the observed fluxes and the lines represent the RWIW models. 
The parameters of these models are listed in Table \ref{rob_parm}.
}
\label{SED_fits}
\end {figure*}

\begin {figure*}
\includegraphics[height=10.0cm]{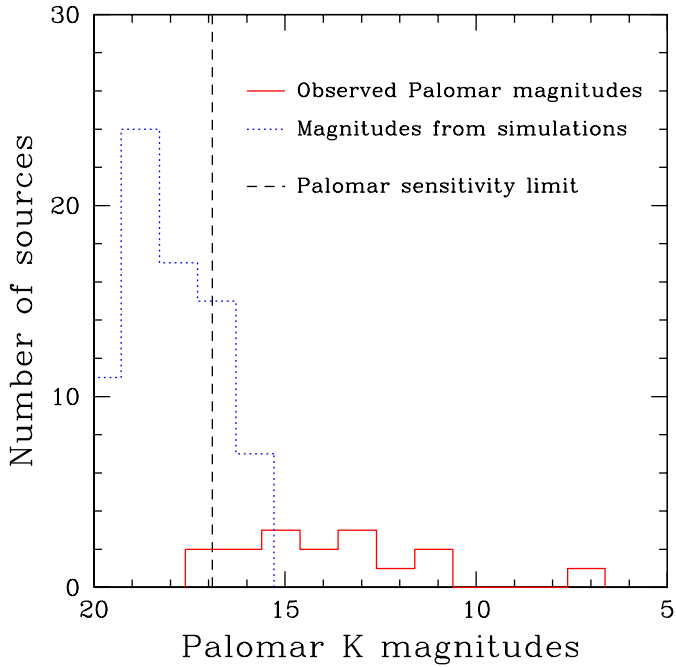}
\caption{Plot of magnitude distribution in the K$_s$ band from the simulations 
and observations (Palomar). The solid line represents the magnitudes 
of YSOs within the `region of interest' (Table \ref{ysolist}). The 
dotted line represents the apparent magnitudes from one 
run of model (general case of constant star formation rate with ages between 
0.01 and 1 Myr) simulations. The dashed line denotes the Palomar 
sensitivity limit.
}
\label{fluxd}
\end {figure*}

\begin {figure*}
\includegraphics[height=10.0cm]{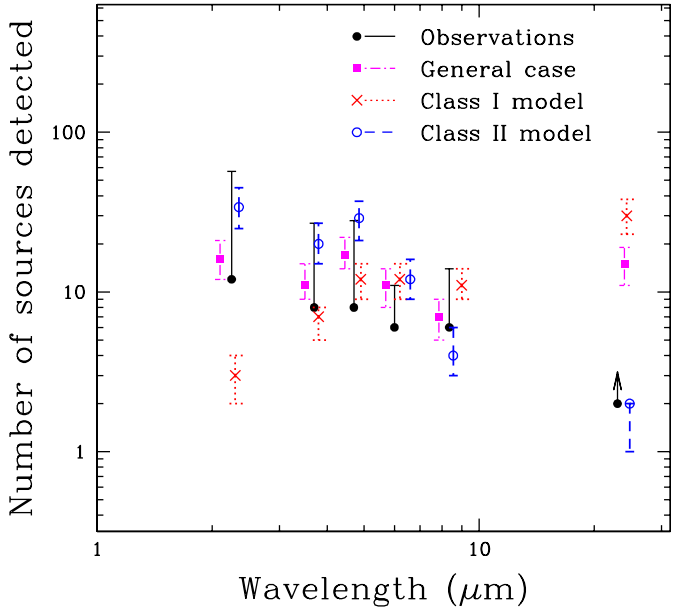}
\caption{Plot of the detected number of sources as a function of wavelength 
for cluster model simulations with observations. The filled circles and 
solid lines represent the observations and error bars, respectively. The 
observation at 24~$\mu$m represents a lower-limit. The solid 
squares represent median values of the general case of simulations 
incorporating Class I as well as Class II objects for the general case of 
constant star formation rate with ages between 0.01 and 1 
Myr and the dot-dashed line represent the quartile values on either side of 
the median. The cross symbols and dotted line represent the Class I model 
simulations for an age of 0.5 Myr. The  open circles and dashed line denote 
Class II model simulations for 1 Myr. The wavelengths for different cases
have been slightly shifted for better viewing. Further details can be found in the text.
}
\label{comp}
\end {figure*}

\begin {figure*}
\includegraphics[height=10.0cm]{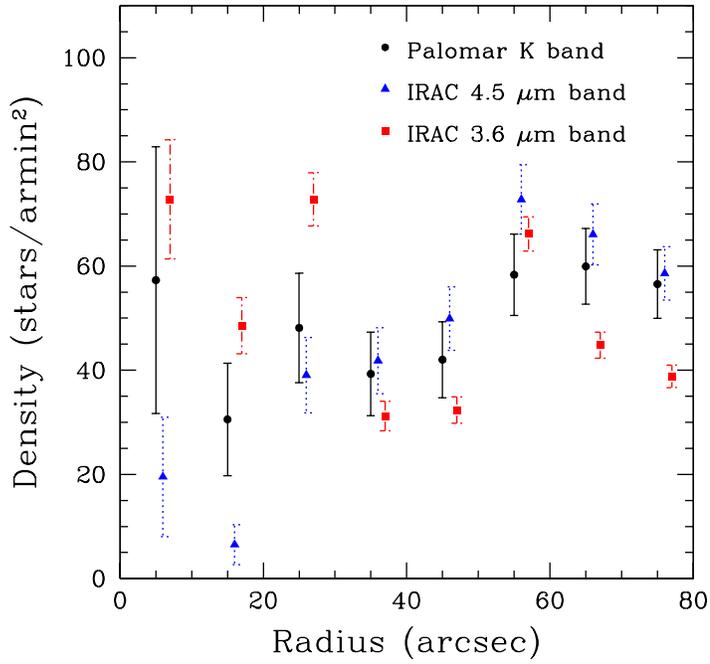}
\caption{ The clustering indicator (radial profile) centred on the bright NIR 
source in IRAS 18511, constructed using the Palomar K$_s$ (solid circles), 
IRAC 4.5~$\mu$m (solid triangles) and IRAC 8.0~$\mu$m (solid squares) band 
fluxes of the sources in the region around IRAS 18511. The star 
densities derived for the IRAC2 and IRAC4 bands have been scaled by 
factors of 1.7 and 6.4, respectively for suitable comparison with the radial 
profile in the K$_s$ band. The radii for 
different cases have been slightly shifted for better viewing.
}
\label{radprof}
\end {figure*}

\clearpage
\appendix

\section{Photometry of Spitzer images}

IRAS 18511 consists of a very bright source and a small group of stars near 
it. We, therefore,  
first explored our photometry techniques on a field consisting of isolated 
sources (which are not saturated) with good signal-to-noise. The following 
methods were used.

\begin{enumerate}
\item{APEX - The Spitzer Astronomical Point Source EXtraction (APEX) is a 
software provided by the Spitzer Science Centre for point source 
extraction and 
photometry. We have used the Linux version MOPEX 030106 for our study. In this 
software, the sources are first detected using non-linear matched filtering and 
image segmentation. Subsequent point source extraction is performed by fitting 
the Point Response Function (PRF) to the detected sources 
\citep{2004AAS...20515312M}. More details can be found in the APEX 
Manual. For PRF fitting, the PRF has been taken from the Spitzer Science Center
 (SSC) webpage. In addition, aperture photometry can also be carried out. We 
have compared the fluxes from apertures 
of radii 6\arcsec, 9\arcsec\ and 12\arcsec.}
\item{DAOPHOT - We have used the IDL version of the IRAF-DAOPHOT package in 
order to carry out source extraction and aperture photometry with an aperture 
radius of 6\arcsec\ and background annulus of 12\arcsec. Appropriate aperture 
corrections have been applied.}
\item{SExtractor - Source Extractor is a program which can be used to extract 
sources from moderately crowded fields. The background in this package is, 
however, determined from the full image and the objects above a given threshold
 are detected and de-blended. For photometry, we have used an aperture radius 
of 6\arcsec.}
\item{AIPS/GILDAS - We have also used the AIPS/GILDAS software to compute the 
integrated flux down to 10\% of the peak value after subtracting the 
background. Appropriate 
aperture corrections have been applied depending on the size of the 
integration area.}
\end{enumerate}

All the above methods did a reasonably good task of source extraction and 
photometry. The 
values of fluxes of the isolated bright sources obtained from the methods 
mentioned above were compared. The values were found to be within 5\% of each 
other. We then proceeded to extract and carry out photometry of the sources 
in the region around IRAS 18511 which has bright saturated sources located 
close to each other as well as some diffuse 
emission. The saturated pixels were replaced by an average value of the 
neighbouring non-saturated pixels. For this region, we find the following:

\begin{itemize}
\item{The source extraction by SEXTRACTOR is better than that by the PRF 
fitting methods, APEX and DAOPHOT.}
\item{The aperture photometry methods yield consistent results within 5\% of 
the flux values.}
\item{The PRF fitting method of APEX finds quite a large number of sources 
near the bright source and the flux is distributed among them.}
\item{The flux values obtained by AIPS/GILDAS which integrates fluxes in 
circular areas down to 10\% of the peak brightness after subtracting 
the background gives consistent results with those from aperture 
photometry.}
\end{itemize}

\section{Description of the cluster simulations}

For a given IMF (Scalo or Salpeter or Kroupa), the cluster members are 
randomly selected by using a Monte-Carlo 
method. The lower mass limit is taken to be 0.1 \Msol. An age (star formation 
hisory) is assigned to each cluster member according to either of the two 
prescriptions:
(a) coeval formation - every cluster member has the same age, or (b) uniform 
formation rate - an age is assigned to every cluster member which falls 
randomly between the ages representing the start and finish of the 
star formation process. The luminosity and effective temperature 
of every cluster member having a certain mass and age is determined using the 
pre-main sequence tracks of \citet{1999ApJ...525..772P}. Because of 
uncertainties in the early pre-main sequence evolution of stars, we 
have assumed a lower limit on the age. The pre main sequence track 
corresponding to 0.5 Myr is applied even to objects with ages less than 0.5 
Myr. For ages of a given mass (as well as masses for a given age) greater than 
that
for the pre-main sequence tracks, the luminosities and effective temperatures
have been obtained by assuming them to be lying on the tracks of zero-age main
sequence (ZAMS) stars. For such Class II objects, it is assumed that although 
the central source has reached the ZAMS stage, there is a remnant disk 
around it. The cluster members are randomly added to the cluster till the 
total luminosity reaches the bolometric luminosity of the cluster.
It is important to note that the the luminosity of each cluster member 
includes luminosities from the central object (photosphere for the Class II 
objects) as well as from the circumstellar 
material (disk for Class II and envelope for Class I objects).

The cluster members are now placed at random locations in a homogeneous 
spherical cloud of gas of assumed mass and size (derived from the 
sub-millimetre/millimetre 
maps). In other words, the distribution of the cluster members follows the gas 
distribution. Thus, there is a column of gas in front of every embedded cluster 
member which determines its extinction. Therefore, different cluster members 
suffer different amounts of extinctions based on their positions within the 
cloud of gas. An additional visual extinction of 7 magnitudes (due to ISM) 
 has been added to the extinctions. Taking the total extinction into 
consideration, the apparent magnitude of every cluster 
member is determined in each band and is considered detected if the object is 
brighter than the sensitivity limit in that band. The sensitivities in various 
bands as well as the extinction laws
used are listed in Table \ref{assump}. The near infrared extinction laws have 
been taken from \citet{1985ApJ...288..618R}. For the Spitzer-IRAC bands, we 
have used the average extinction value obtained by \citet{2005ApJ...619..931I}.
 At 24~$\mu$m, we have used an extrapolated extinction value. 

Every cluster member is classified as Class I or Class II. Sources 
younger than 0.5 Myr are considered as Class I and those older than 0.5 Myr 
are considered to be of Class II type. We have taken the age of 0.5 
Myr between the Class I and more evolved objects as the pre-main sequence 
evolution models show uncertainities below the age of 1 Myr 
\citep{2002A&A...382..563B}. For Class I sources, the flux in various bands 
has been determined based on 
the spectrum of the well-studied Class I type low mass protostar, L1551-IRS5.
\citet{1983ApJ...265..877C} list the magnitudes of L1551-IRS5 in 11 bands 
ranging from 1.2~$\mu$m to 19~$\mu$m. The ratio of the absolute flux 
(corrected for distance of L1551) to the luminosity of L1551-IRS5 ($\sim35$ 
\Lsol) has been used as a scaling factor to obtain the fluxes of Class I type 
objects in the simulations corresponding to the luminosities of these objects. 
The ratios of fluxes (in various bands) to the total luminosity are tabulated 
in Table \ref{assump}.

In order to determine the emission from pre-main sequence objects
 in the Class II phase, we have used the median spectrum of Class II type 
objects in the Taurus star forming region 
\citep[Table 6 of][]{2006ApJS..165..568F}. This median spectrum has been 
constructed using fluxes from bands including the near infrared (JHK$_s$) and 
Spitzer-IRAC (3.6, 4.5, 5.8 and 8.0~$\mu$m) bands. This median spectrum 
implies flux 
ratios (flux in various bands with respect to the flux in the H-band) which 
are given in Table \ref{assump}. The normalization has been carried out with 
respect to the flux values in H (1.65~$\mu$m) band since the H-band flux is 
photospheric for most Class II objects \citep{2006ApJS..165..568F}. 
As the Class I and Class II spectra are of low-mass objects, we have 
used these in the model simulations for objects of all masses. Although the 
pre-main sequence evolution of massive stars is not clearly understood, we use 
these Class I and Class II spectra since fits from RWIW 
models (Sect. 3.6.1) indicate that the SEDs of cluster members of IRAS 18511 
are well-fit by models comprising massive central objects. The Class 
I and Class II type spectral energy distributions, used in the model, are 
shown in Fig \ref{sim_sedAv}. These spectra have been normalised with respect 
to the H (1.65~$\mu$m) band. The Spitzer colours of Class I and Class II 
SEDs considered for modelling are shown in Fig \ref{ccd}.

\begin{table*}
\caption{Extinction laws and various other constants used in the simulations. 
F$_{\lambda}$/F$_H$ represents the ratio of flux in each band with respect to 
the flux in H band assumed for Class II sources in the simulations based on 
the median spectrum of Class II sources by \citet{2006ApJS..165..568F} and 
$\lambda$/L represents the ratio of flux in each band to the 
luminosity of L1551-IRS5 \citep{1983ApJ...265..877C} used to estimate the 
fluxes of Class I sources in the simulations.}
\label{assump}
\begin{tabular}{c c c c c} \hline \hline
$\lambda$ & A$_{\lambda}$/A$_V$ & F$_{\lambda}$/F$_H$ & $\lambda$/L & Sensitivity limit \\ \hline
$\mu$m    &                     &                     & ($\mu$m/\Lsol) & \\ \hline
2.2 & 0.112  & 0.979  & 0.685  &  16.9 \\
3.5 & 0.063  & 0.787  & 2.107  & 14.5 \\
4.6 & 0.048  & 0.714  & 9.880  & 14.0 \\
5.8 & 0.048  & 0.645  & 14.65  & 13.0 \\
8   & 0.048  & 0.654  & 27.51  & 11.5 \\
24  & 0.005  & 1.326  & 495.0  & 6.8    \\
\hline
\end{tabular}
\end{table*}

\end{document}